\documentclass[aps,prl,twocolumn,superscriptaddress,amsmath,amssymb,floatfix]{revtex4-2}

\usepackage{graphicx}
\usepackage{tikz}
\usepackage{xcolor}
\usepackage[colorlinks=true,linkcolor=blue!60!black,citecolor=blue!60!black,urlcolor=blue!60!black]{hyperref}
\usepackage{orcidlink}
\usepackage[normalem]{ulem}

\newcommand{\sg}{\sigma}

\newcommand{\up}{\uparrow}
\newcommand{\dn}{\downarrow}
\newcommand{\ket}[1]{\left|#1 \right.\rangle}
\newcommand{\bra}[1]{\langle#1|}
\newcommand{\avg}[1]{\langle#1\rangle}
\newcommand{\R}{\mathcal{R}}
\newcommand{\et}{\tilde e}

\newcommand{\phead}[1]{\textit{#1}.---}

\usepackage{pdfpages}

\makeatletter
\AtBeginDocument{\let\LS@rot\@undefined}
\makeatother
\begin{document}

\newcommand{\unige}{
  Department of Theoretical Physics, University of Geneva,
  24 quai Ernest-Ansermet, 1211 Gen\`eve, Switzerland
}

\newcommand{\gqc}{
  Geneva Quantum Center, University of Geneva
}

\newcommand{\princeton}{
Department of Electrical and Computer Engineering, Princeton University, Princeton, NJ 08544, USA
}

\title{Long-lived local coherences at high temperature from frozen domain walls}

\author{Sarang Gopalakrishnan\:\orcidlink{0000-0002-7493-7600}}
\affiliation{\princeton}
\author{Romain Vasseur\:\orcidlink{0000-0002-4636-4139}}
\affiliation{\unige}
 \affiliation{\gqc}

\date{\today}

\begin{abstract}
 Local coherences between distinct charge sectors---measured by the single-particle Green's function or the transverse spin autocorrelation function---do not overlap with hydrodynamic modes. At finite temperature, therefore, they decay faster than any power law in generic chaotic systems. We present a mechanism by which these coherences can persist anomalously long, for integrable and chaotic spin chains (including the XXZ and Fredkin spin chains) that obey a quantum group symmetry. We show---using a Mazur bound derived from the quantum-group symmetry, together with locality and a ``non-undershoot'' assumption---that the transverse spin autocorrelation function $C(t) \equiv \langle \sigma^-_x(t) \sigma^+_x(0)\rangle$ can decay at most algebraically: $C(t) \gtrsim 1/t^2$. This bound is not tight: in the large-anisotropy limit of the XXZ spin chain, at infinite temperature, we explicitly compute $C(t) \sim \sqrt{3}/(18 \sqrt{t})$. We also compute this decay rate in the XXZ spin chain for general easy-axis anisotropy, and estimate it to scale as $C(t) \sim t^{-2/z}$ (for transport exponent $z$) in the chaotic models.
\end{abstract}

\maketitle

\phead{Introduction}
A local operator in a chaotic many-body system loses its memory quickly unless it overlaps with hydrodynamic operators---conserved densities, or their powers or derivatives~\cite{ForsterBook,Lux2014,KVH2018,PhysRevX.8.031058}. 
Operators that change the value of a conserved charge---such as the creation operator for fermions, or the spin raising operator for $U(1)$-symmetric spin systems---are manifestly orthogonal to all hydrodynamic operators. 
Therefore one expects these ``transverse'' operators to be immune to hydrodynamic long-time tails.
Physically, transverse operators create local coherences between distinct charge sectors, which the rest of the system rapidly ``measures'' and dephases. 
In high dimensions, transverse operators decay exponentially. In one dimension this decay is \emph{stretched-exponential}~\cite{MJKG2025, MJG2026}: coherences can persist anomalously long in low-density regions where dephasing is slow. 
Slow dephasing is of interest both because transverse correlators are routinely measured in experiment (e.g., via ARPES), and because they offer a readily measurable signature of quantum coherence even at high temperatures in chaotic many-body systems. 
In practice, stretched exponentials still decay rapidly, so it is natural to ask what minimal additional structure one needs to slow the decay even further to a power law. A recent paper~\cite{DGWZZ} proposed (based on numerics) that a \emph{quantum group} symmetry might accomplish this, even in chaotic systems. 

\begin{figure}[b]
\centering
\resizebox{\columnwidth}{!}{\begin{tikzpicture}[font=\sffamily\small]
\definecolor{cA}{HTML}{2a78d6}\definecolor{cR}{HTML}{d2383a}\definecolor{ink2}{HTML}{52514e}\definecolor{band}{HTML}{fbe7b5}
\fill[band,rounded corners=1.5pt] (3.615,-0.280) rectangle (3.885,2.831);
\draw[black!35,densely dashed,line width=0.3pt] (0.150,2.121) -- (0.150,1.050);
\draw[black,line width=0.55pt,-{stealth}] (0.150,2.771) -- (0.150,2.541);
\draw[cR,line width=0.55pt,-{stealth}] (0.150,2.411) -- (0.150,2.181);
\draw[black!35,densely dashed,line width=0.3pt] (0.450,2.121) -- (0.450,1.050);
\draw[black,line width=0.55pt,-{stealth}] (0.450,2.541) -- (0.450,2.771);
\draw[cR,line width=0.55pt,-{stealth}] (0.450,2.181) -- (0.450,2.411);
\draw[black!35,densely dashed,line width=0.3pt] (0.750,2.121) -- (0.750,1.050);
\draw[black,line width=0.55pt,-{stealth}] (0.750,2.771) -- (0.750,2.541);
\draw[cR,line width=0.55pt,-{stealth}] (0.750,2.411) -- (0.750,2.181);
\draw[black!35,densely dashed,line width=0.3pt] (1.050,2.121) -- (1.050,0.770);
\draw[black,line width=0.55pt,-{stealth}] (1.050,2.771) -- (1.050,2.541);
\draw[cR,line width=0.55pt,-{stealth}] (1.050,2.411) -- (1.050,2.181);
\draw[black!35,densely dashed,line width=0.3pt] (1.350,2.121) -- (1.350,0.770);
\draw[black,line width=0.55pt,-{stealth}] (1.350,2.541) -- (1.350,2.771);
\draw[cR,line width=0.55pt,-{stealth}] (1.350,2.181) -- (1.350,2.411);
\draw[black!35,densely dashed,line width=0.3pt] (1.650,2.121) -- (1.650,0.770);
\draw[black,line width=0.55pt,-{stealth}] (1.650,2.771) -- (1.650,2.541);
\draw[cR,line width=0.55pt,-{stealth}] (1.650,2.411) -- (1.650,2.181);
\draw[black!35,densely dashed,line width=0.3pt] (1.950,2.121) -- (1.950,0.770);
\draw[black,line width=0.55pt,-{stealth}] (1.950,2.541) -- (1.950,2.771);
\draw[cR,line width=0.55pt,-{stealth}] (1.950,2.181) -- (1.950,2.411);
\draw[black!35,densely dashed,line width=0.3pt] (2.250,2.121) -- (2.250,0.770);
\draw[black,line width=0.55pt,-{stealth}] (2.250,2.771) -- (2.250,2.541);
\draw[cR,line width=0.55pt,-{stealth}] (2.250,2.411) -- (2.250,2.181);
\draw[black!35,densely dashed,line width=0.3pt] (2.550,2.121) -- (2.550,0.490);
\draw[black,line width=0.55pt,-{stealth}] (2.550,2.771) -- (2.550,2.541);
\draw[cR,line width=0.55pt,-{stealth}] (2.550,2.411) -- (2.550,2.181);
\draw[black!35,densely dashed,line width=0.3pt] (2.850,2.121) -- (2.850,0.490);
\draw[black,line width=0.55pt,-{stealth}] (2.850,2.541) -- (2.850,2.771);
\draw[cR,line width=0.55pt,-{stealth}] (2.850,2.181) -- (2.850,2.411);
\draw[black!35,densely dashed,line width=0.3pt] (3.150,2.121) -- (3.150,0.490);
\draw[black,line width=0.55pt,-{stealth}] (3.150,2.771) -- (3.150,2.541);
\draw[cR,line width=0.55pt,-{stealth}] (3.150,2.411) -- (3.150,2.181);
\draw[black!35,densely dashed,line width=0.3pt] (3.450,2.121) -- (3.450,0.210);
\draw[black,line width=0.55pt,-{stealth}] (3.450,2.771) -- (3.450,2.541);
\draw[cR,line width=0.55pt,-{stealth}] (3.450,2.411) -- (3.450,2.181);
\draw[black!35,densely dashed,line width=0.3pt] (3.750,2.121) -- (3.750,0.210);
\draw[black,line width=0.55pt,-{stealth}] (3.750,2.771) -- (3.750,2.541);
\draw[cR,line width=0.55pt,-{stealth}] (3.750,2.181) -- (3.750,2.411);
\draw[black!35,densely dashed,line width=0.3pt] (4.050,2.121) -- (4.050,0.490);
\draw[black,line width=0.55pt,-{stealth}] (4.050,2.541) -- (4.050,2.771);
\draw[cR,line width=0.55pt,-{stealth}] (4.050,2.181) -- (4.050,2.411);
\draw[black!35,densely dashed,line width=0.3pt] (4.350,2.121) -- (4.350,0.770);
\draw[black,line width=0.55pt,-{stealth}] (4.350,2.541) -- (4.350,2.771);
\draw[cR,line width=0.55pt,-{stealth}] (4.350,2.181) -- (4.350,2.411);
\draw[black!35,densely dashed,line width=0.3pt] (4.650,2.121) -- (4.650,0.770);
\draw[black,line width=0.55pt,-{stealth}] (4.650,2.771) -- (4.650,2.541);
\draw[cR,line width=0.55pt,-{stealth}] (4.650,2.411) -- (4.650,2.181);
\draw[black!35,densely dashed,line width=0.3pt] (4.950,2.121) -- (4.950,0.770);
\draw[black,line width=0.55pt,-{stealth}] (4.950,2.541) -- (4.950,2.771);
\draw[cR,line width=0.55pt,-{stealth}] (4.950,2.181) -- (4.950,2.411);
\draw[black!35,densely dashed,line width=0.3pt] (5.250,2.121) -- (5.250,1.050);
\draw[black,line width=0.55pt,-{stealth}] (5.250,2.541) -- (5.250,2.771);
\draw[cR,line width=0.55pt,-{stealth}] (5.250,2.181) -- (5.250,2.411);
\draw[black!35,densely dashed,line width=0.3pt] (5.550,2.121) -- (5.550,1.330);
\draw[black,line width=0.55pt,-{stealth}] (5.550,2.541) -- (5.550,2.771);
\draw[cR,line width=0.55pt,-{stealth}] (5.550,2.181) -- (5.550,2.411);
\draw[black!35,densely dashed,line width=0.3pt] (5.850,2.121) -- (5.850,1.330);
\draw[black,line width=0.55pt,-{stealth}] (5.850,2.771) -- (5.850,2.541);
\draw[cR,line width=0.55pt,-{stealth}] (5.850,2.411) -- (5.850,2.181);
\draw[black!35,densely dashed,line width=0.3pt] (6.150,2.121) -- (6.150,1.330);
\draw[black,line width=0.55pt,-{stealth}] (6.150,2.541) -- (6.150,2.771);
\draw[cR,line width=0.55pt,-{stealth}] (6.150,2.181) -- (6.150,2.411);
\draw[black!35,densely dashed,line width=0.3pt] (6.450,2.121) -- (6.450,1.610);
\draw[black,line width=0.55pt,-{stealth}] (6.450,2.541) -- (6.450,2.771);
\draw[cR,line width=0.55pt,-{stealth}] (6.450,2.181) -- (6.450,2.411);
\draw[black!35,densely dashed,line width=0.3pt] (6.750,2.121) -- (6.750,1.610);
\draw[black,line width=0.55pt,-{stealth}] (6.750,2.771) -- (6.750,2.541);
\draw[cR,line width=0.55pt,-{stealth}] (6.750,2.411) -- (6.750,2.181);
\draw[black!35,densely dashed,line width=0.3pt] (7.050,2.121) -- (7.050,1.610);
\draw[black,line width=0.55pt,-{stealth}] (7.050,2.541) -- (7.050,2.771);
\draw[cR,line width=0.55pt,-{stealth}] (7.050,2.181) -- (7.050,2.411);
\draw[black!35,densely dashed,line width=0.3pt] (7.350,2.121) -- (7.350,1.890);
\draw[black,line width=0.55pt,-{stealth}] (7.350,2.541) -- (7.350,2.771);
\draw[cR,line width=0.55pt,-{stealth}] (7.350,2.181) -- (7.350,2.411);
\node[black,anchor=east] at (-0.02,2.656) {$\langle A|$};
\node[cR,anchor=east] at (-0.02,2.296) {$|B_A\rangle$};
\draw[ink2,|-|] (0.30,-0.73) -- node[below=-1pt]{window $W$} (7.20,-0.73);
\draw[ink2,densely dotted] (0.30,-0.73) -- (0.30,0.84); \draw[ink2,densely dotted] (7.20,-0.73) -- (7.20,1.12);
\draw[cR,line width=0.9pt] (3.60,0.00) -- (3.90,0.28) -- (4.20,0.56) -- (4.50,0.84) -- (4.80,0.56) -- (5.10,0.84) -- (5.40,1.12) -- (5.70,1.40) -- (6.00,1.12) -- (6.30,1.40) -- (6.60,1.68) -- (6.90,1.40) -- (7.20,1.68) -- (7.50,1.96);
\draw[black,line width=0.9pt] (0.00,1.12) -- (0.30,0.84) -- (0.60,1.12) -- (0.90,0.84) -- (1.20,0.56) -- (1.50,0.84) -- (1.80,0.56) -- (2.10,0.84) -- (2.40,0.56) -- (2.70,0.28) -- (3.00,0.56) -- (3.30,0.28) -- (3.60,0.00) -- (3.90,-0.28) -- (4.20,0.00) -- (4.50,0.28) -- (4.80,0.00) -- (5.10,0.28) -- (5.40,0.56) -- (5.70,0.84) -- (6.00,0.56) -- (6.30,0.84) -- (6.60,1.12) -- (6.90,0.84) -- (7.20,1.12) -- (7.50,1.40);
\fill[cR] (3.90,0.28) circle (0.7pt);
\fill[cR] (4.20,0.56) circle (0.7pt);
\fill[cR] (4.50,0.84) circle (0.7pt);
\fill[cR] (4.80,0.56) circle (0.7pt);
\fill[cR] (5.10,0.84) circle (0.7pt);
\fill[cR] (5.40,1.12) circle (0.7pt);
\fill[cR] (5.70,1.40) circle (0.7pt);
\fill[cR] (6.00,1.12) circle (0.7pt);
\fill[cR] (6.30,1.40) circle (0.7pt);
\fill[cR] (6.60,1.68) circle (0.7pt);
\fill[cR] (6.90,1.40) circle (0.7pt);
\fill[cR] (7.20,1.68) circle (0.7pt);
\fill[cR] (7.50,1.96) circle (0.7pt);
\fill[black] (0.00,1.12) circle (0.7pt);
\fill[black] (0.30,0.84) circle (0.7pt);
\fill[black] (0.60,1.12) circle (0.7pt);
\fill[black] (0.90,0.84) circle (0.7pt);
\fill[black] (1.20,0.56) circle (0.7pt);
\fill[black] (1.50,0.84) circle (0.7pt);
\fill[black] (1.80,0.56) circle (0.7pt);
\fill[black] (2.10,0.84) circle (0.7pt);
\fill[black] (2.40,0.56) circle (0.7pt);
\fill[black] (2.70,0.28) circle (0.7pt);
\fill[black] (3.00,0.56) circle (0.7pt);
\fill[black] (3.30,0.28) circle (0.7pt);
\fill[black] (3.60,0.00) circle (0.7pt);
\fill[black] (3.90,-0.28) circle (0.7pt);
\fill[black] (4.20,0.00) circle (0.7pt);
\fill[black] (4.50,0.28) circle (0.7pt);
\fill[black] (4.80,0.00) circle (0.7pt);
\fill[black] (5.10,0.28) circle (0.7pt);
\fill[black] (5.40,0.56) circle (0.7pt);
\fill[black] (5.70,0.84) circle (0.7pt);
\fill[black] (6.00,0.56) circle (0.7pt);
\fill[black] (6.30,0.84) circle (0.7pt);
\fill[black] (6.60,1.12) circle (0.7pt);
\fill[black] (6.90,0.84) circle (0.7pt);
\fill[black] (7.20,1.12) circle (0.7pt);
\fill[black] (7.50,1.40) circle (0.7pt);
\node[inner sep=0pt] at (3.798,0.000) {$x$};
\fill[cA] (3.90,-0.28) circle (1.8pt);
\node[cA,below] at (3.90,-0.30) {valley};
\node[ink2,anchor=east,align=right] (mob) at (2.63,-0.20) {mobile\\[-2.5pt]excitation};
\draw[ink2,->] (mob.east) to[out=0,in=-100] (2.88,0.34);
\end{tikzpicture}}
\caption{{\bf Mechanism in the height representation} $h(i)=\sum_{j<i}2S^z_j$. The valley is the unique lowest point of the window $W$. Flipping the down step $x$ into the valley (dashed) raises the path to its right by two units and moves the valley by one site, creating no excitation. A mobile excitation (here a lone up step in the falling run) crosses the valley and displaces it by two sites in both branches alike. The correlator of this configuration is the probability that the valley is back at $x$.}
\label{fig:schematic}
\end{figure}
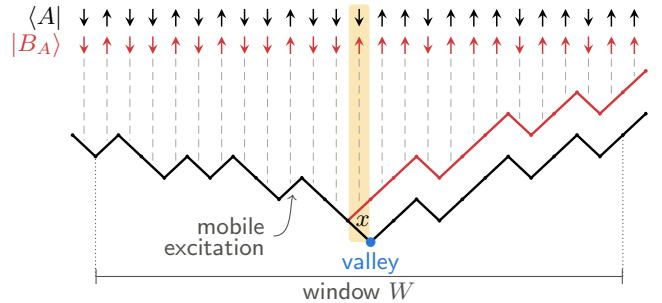

In this paper we explain the mechanism behind this slow decay of the infinite-temperature transverse spin correlator $C(t) \equiv \langle \sigma^-_x(t) \sigma^+_x(0) \rangle$ in quantum-group models~\cite{PasquierSaleur1990}. This mechanism is most naturally described in the height representation (Fig.~\ref{fig:schematic}), where each $\downarrow (\uparrow)$ spin maps to a down (up) step in the height field. Heights live on bonds in the original lattice; $h(x)$ is the height
on the bond immediately to the left of site $x$. Under generic quantum evolution, height minima have no special status, but in the presence of quantum group symmetries they host quantum coherence. Specifically, for any window $W$ of the spin chain, the Hamiltonian $H_W$ restricted to $W$ admits an operator $\mathcal{R}_W$ local to the window $W$ that (i) commutes with the Hamiltonian $H_W$ restricted to the window, and (ii) acts on a product state $|c_x\rangle$ with a local minimum at $x$ as $\mathcal{R}_W |c_x\rangle \approx \sigma^+_x |c_x\rangle$. The action of $\sigma^+_x$ is temporarily invisible to the other degrees of freedom, so they cannot dephase the coherence it creates. This invisibility lasts as long as the minimum in question is the deepest one in its light cone; while it lasts, $C(t)$ is simply related to the return probability of the minimum. In the infinite temperature state of a large system, the minimum of a region of length $\ell$ is typically at least $\ell$ away from any deeper minimum, so the spin-flip remains invisible for a timescale $\agt \ell$. Together with the power-law decaying return probability of the minimum, this property suffices to make $C(t)$ algebraically long-lived. 

Our main results are threefold. First, we present a general argument, based purely on the quantum-group symmetry, that $C(t)$ decays at most as fast as $1/t^2$. This argument is rigorous except for a physically motivated (and numerically verified) ``non-undershoot'' assumption. 
Next, we present a heuristic theory, decomposing the survival of transverse correlators into two components, the survival and transport of local minima. We support this theory by an exact calculation for the large-anisotropy limit of the XXZ spin chain~\cite{ZF1,ZBBF,Pozsgay}. (Ref.~\cite{JGE2009} previously studied $C(t)$ at low temperatures in this model; our high-temperature results involve qualitatively different considerations.)
For this model, $C(t) \sim \sqrt{3}/(18 \sqrt{t})$ in the late time limit. 
We extend our analysis to the case of finite easy-axis anisotropy, as well as to the chaotic Fredkin and quantum-group spin chains~\cite{SalbergerKorepin,MSGV2025}. (To our knowledge, the quantum-group structure of the Fredkin spin chain has not been discussed previously; see, however, Refs.~\cite{Salberger2021thesis,Salberger2018Motzkin,Hao2023Motzkin}.)
Finally, we comment on magnetized ensembles, where the integrable models retain long-lived coherences but the chaotic models cross over to generic stretched exponential decay. 

Two points are worth emphasizing. First, the slow decay of transverse correlators is \emph{not} generic for integrable systems: indeed, $C(t)$ decays exponentially (or faster) even in the XXZ model for easy-plane anisotropy~\cite{BJ,CapelPerk,GKSS}. Second, the quantum group symmetry cannot be associated with a hydrodynamic mode. One way to see this is that the power-law decay of $C(t)$ is restricted to half-filling. By contrast, if the raising operator overlapped with a hydrodynamic mode (as it does for isotropic spin chains), it would decay algebraically for all densities. Hence the long-lived coherences are not just long-time tails. 

\phead{Models}
We consider local spin-$\frac{1}{2}$ chains that obey a certain quantum group $U_q(sl_2)$ symmetry~\cite{PasquierSaleur1990,ABBBQ1987}, a ``$q$-deformed'' version of $SU(2)$, which are built from the Temperley--Lieb (TL) generators $e_m=(q+q^{-1})\ket{\psi_q}\bra{\psi_q}$, $q>1$, scaled projectors on the $q$-singlet $\ket{\psi_q}\propto q^{1/2}\ket{\up\dn}-q^{-1/2}\ket{\dn\up}$ of the bond $(m,m+1)$. They satisfy the TL algebra  $e_m^2=(q+q^{-1})e_m$, $e_me_{m\pm1}e_m=e_m$ and $[e_n,e_m]=0$ for $\left| n-m\right|>1$~\footnote{In terms of Pauli operators, $e_m=-\tfrac12\big(\sg^x_m\sg^x_{m+1}+\sg^y_m\sg^y_{m+1}\big)-\tfrac\Delta2\,\sg^z_m\sg^z_{m+1}+\tfrac{q-q^{-1}}{4}\big(\sg^z_m-\sg^z_{m+1}\big)+\tfrac\Delta2$}. Any Hamiltonian that can be written in terms of the $e_m$'s only commutes with the quantum group action. Our results apply to all of these, but for definiteness (and to connect with previous work) we focus on the family of three-site Hamiltonians
\begin{equation}
H(\beta) =\sum_j\Big[-(e_j+e_{j+1})+\beta\,(e_je_{j+1}+e_{j+1}e_j)\Big].
\label{eq:HQG}
\end{equation}
Eq.~\eqref{eq:HQG} encompasses many standard quantum spin chains. 
For $\beta = 0$, this is the anisotropic XXZ spin chain $H = J \sum_i S^x_i S^x_{i+1} + S^y_i S^y_{i+1} + \Delta S^z_i S^z_{i+1}$ with $S_k^z=\sg_k^z/2$ and $J=4$ (up to boundary conditions), for $\Delta = \tfrac12(q+q^{-1})>1$. For generic $\beta$ it is the chaotic chain studied numerically in Ref.~\cite{DGWZZ}.
%
As $q\to\infty$ the generators become diagonal, $e_m=q\, n^\up_mn^\dn_{m+1} +O(1)$ projecting onto a peak $\up\dn$ of the height profile, while their products retain a hopping of order $q$ that generate the so-called Fredkin Hamiltonian $H_{\rm F}=\sum_j[n^\up_jP^S_{j+1,j+2}+P^S_{j,j+1}n^\dn_{j+2}]$~\cite{SalbergerKorepin,DellAnna,SUZKKM,MovassaghShor}, where $P^S$ is the singlet projector. In that limit, $H/q \to2\beta H_{\rm F}-(1+\beta)\sum_j\big(n^\up_j n^\dn_{j+1}+n^\up_{j+1}n^\dn_{j+2}\big)$ where the Fredkin part moves a local peak over an adjacent spin, and the second term is an energy per peak. For $\beta = -1$, in particular, $H(\beta)$ reduces to the Fredkin spin chain as $q \to \infty$ (since peak energies are not penalized). Meanwhile, for $\beta = 0$, the dynamics fragments~\cite{SalaFrag,KHN2020} into sectors with fixed peak number, and we recover the so-called folded XXZ spin chain~\cite{ZF1,ZBBF,Pozsgay}, $H_{\rm f}=\frac J2\sum_j\frac{1+\sg^z_{j-1}\sg^z_{j+2}}{2}\big(\sg^+_j\sg^-_{j+1}+{\rm h.c.}\big)$ which describes the $\Delta\to\infty$ limit of the integrable XXZ chain. In this model, only lone spins whose neighbors are equal and opposite to them can move; all other spins are frozen.

\begin{figure*}[t]
\centering
\includegraphics[width=\textwidth]{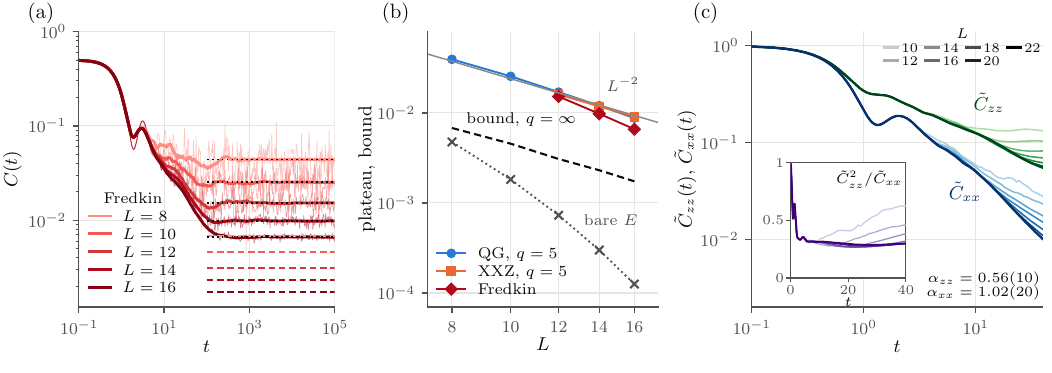}
\caption{{\bf Mazur bound and chaotic chains}. (a) Transverse autocorrelator at the center of open Fredkin chains of $L=8$--$16$ sites from full diagonalization (thin lines) and its time average $\frac1t\int_t^{2t}C dt$ (thick, drawn at $\sqrt2\,t$), with the infinite-time average (dotted) and the Mazur bound (dashed). (b) Infinite-time average against $L$ for the quantum-group chain at $q=5$ and $\beta=0.32$, the integrable chain ($\beta=0$), and the Fredkin chain; the line is $\propto L^{-2}$, and the crosses are the bound with the bare charge $E$. The dashed line is the Mazur bound Eq.~\eqref{eq:plateau}. (c) Decay of moving-averaged transverse ($C_{xx} \sim t^{-\alpha_{xx}}$) and longitudinal ($C_{zz} \sim t^{-\alpha_{zz}}$) correlators for the Fredkin spin chain at various sizes, computed using exact diagonalization ($L \leq 14$) and dynamical typicality ($L > 14$). Inset shows that the ratio of $\tilde C_{zz}^2/\tilde{C}_{xx}$ approaches a constant, as predicted by Eq.~\eqref{eq:structure}. Although neither correlator has reached its asymptotic power law at these sizes, the ratio $\tilde C_{zz}^2/\tilde C_{xx}$ is already constant. } 
\label{fig:mazur}
\end{figure*}

\phead{Finite-size Mazur bound}
Every term of Eq.~\eqref{eq:HQG} commutes with the generators of $U_q(sl_2)$, including the raising generator
\begin{equation}
E=\sum_{i}
q^{-\sum_{k<i} S_k^z}\sg_i^+
q^{\sum_{k>i} S_k^z}.
\label{eq:E}
\end{equation}
For a spin configuration $c=\lbrace s_i\rbrace$ this gives $E\ket c=q^{M+1/2}\sum_{i: s_i=\downarrow}q^{-h_c(i)}\ket{\sg_i^+c}$. Here $M=\sum_k S^z_k$ is the total magnetization, $h_c(i)=2 \sum_{k<i} S^z_k$ is the height immediately before site $i$.
Since $E$ \emph{could} in principle overlap with $\sigma^+$ it is a natural candidate for a Mazur bound~\cite{Mazur1969,Suzuki1971}. However, the exponential factors of $q$ strongly weight atypical configuration pairs, such as $\ket{\downarrow^{L/2-1}\uparrow^{L/2+1}}\langle\downarrow^{L/2}\uparrow^{L/2}|$. By contrast, the local operator $\sigma_i^+$ has equal matrix elements for every allowed spin flip. Consequently, its normalized overlap with $E$ is exponentially small in the system size~\cite{DGWZZ}. 

The remedy is the polar decomposition $E=\R|E|$ with $|E|=(E^\dagger E)^{1/2}$ which defines a partial isometry $\R$. Since $H$ is Hermitian and commutes with $E$, it also commutes with $E^\dagger$ and hence with $\R$. In the quantum-group spin basis, $\R$ raises the magnetization within each multiplet with unit amplitude, removing the exponentially varying coefficients of $E$. At fixed magnetization, the eigenspaces of $E^\dagger E$ approach configuration subspaces of fixed depth as $q\to\infty$, and $\R$ tends to the ``crystal raising operator'' $\et$~\cite{SM} introduced by Kashiwara~\cite{Kashiwara1990,Kashiwara1991}. On a product state, $\et$ flips the last unmatched down spin, equivalently the down step that first reaches the global height minimum. It annihilates the state if that minimum is attained at the left edge. Using $\R$ at finite $q$ and its limit $\et$ for the $q=\infty$ chains allows us to formulate the Mazur bound uniformly across the family of models introduced above.

We now consider an open chain on a window of length $L$ with Hamiltonian $H_L$. As noted before, $[\R, H_L] = 0$. For the correlator $C(t)=\avg{\sg^-_x(t)\sg^+_x}$, the infinite-temperature Mazur bound due to $\R$ is~\cite{Mazur1969,Suzuki1971,ZNP1997},
\begin{equation}
\overline{C}\equiv\lim_{T\to\infty}\frac1T\!\int_0^T\!\!C(t)\,dt\;\ge\;\frac{|\avg{\sg^-_x \R}|^2}{\avg{\R^\dagger \R}} .
\label{eq:mazur}
\end{equation}
For even $L$, each quantum-group multiplet contains one highest-weight state and one state with $S^z=0$. Since $\R^\dagger\R$ projects onto states that are not highest weights, $\avg{\R^\dagger\R}=1-\dim\mathcal H_{S^z=0}/2^L=1-2^{-L}\binom{L}{L/2}$. In the $q\to \infty$ limit, the static overlap $\avg{\sg_x^-\et}=p_{L,x}$ counts configurations whose first global height minimum occurs at $x$. At the center of a large window, $p_{L,x}\sim2/(\pi L)$, giving
\begin{equation}
\overline C\ge
\frac{p_{L,x}^2}{1-2^{-L}\binom{L}{L/2}}
\sim\frac{4}{\pi^2L^2}.
\label{eq:plateau}
\end{equation}
At finite $q$, the static overlap is no smaller than its $q=\infty$ value, $\avg{\sg_x^-\R_q}\ge p_{L,x}$~\cite{SM}. Since $\avg{\R_q^\dagger\R_q}$ is $q$-independent, the same Mazur lower bound holds for all $q>1$.
Figure~\ref{fig:mazur}(a) shows the approach of $C(t)$ to the plateau, and Fig.~\ref{fig:mazur}(b) compares the bound with the exact plateau.

\phead{Finite-time bound from locality}
In the thermodynamic limit, the finite-size Mazur bound becomes trivial. Nevertheless, locality allows us to relate it to the finite-\emph{time} decay of correlators. Consider a window of $2\ell+1$ sites centered at $x$, with restricted Hamiltonian $H_W$ and correlator $C_W(t)$. The Lieb--Robinson bound~\cite{LiebRobinson,NachtergaeleSims2006} ensures that $|C(t)-C_W(t)|\le\epsilon$ whenever $\ell\ge vt+\xi\ln(Kt/\epsilon)$, where $v$ is the Lieb--Robinson velocity and $\xi$, $K$ are constants.

The Mazur bound constrains the infinite-time average of $C_W(t)$, rather than its value at each time. To suppress microscopic oscillations while retaining the long-time decay, we average the correlator over the interval from $t$ to $2t$: $\tilde C_W(t)\equiv\frac{1}{t}\int_t^{2t}C_W(\tau)\,d\tau$. On physical grounds, we expect this smoothed correlator to remain above its late-time plateau before boundary effects become appreciable. This ``non-undershoot'' expectation can be formalized and checked numerically~\cite{SM}: our numerics support it well beyond the times required by the locality argument. To compare with the full chain throughout the averaging interval, we choose $\ell\simeq2vt$, with a logarithmic correction ensuring $\epsilon=o(t^{-2})$. The finite-size plateau in Eq.~\eqref{eq:plateau} then gives
\begin{equation}
\tilde C(t)\equiv\frac{1}{t}\int_t^{2t}C(\tau)\,d\tau \;\gtrsim\;
\frac{1}{4 \pi^2v^2t^2}.
\label{eq:lcbound}
\end{equation}
Since $C(\tau)$ must reach at least its average somewhere in each interval $[t,2t]$, there are arbitrarily late times at which it is at least of order $t^{-2}$. In particular, $t^2C(t)$ cannot converge to zero.

\phead{Chaotic case: hydrodynamic argument} 
Eq.~\eqref{eq:lcbound} can be shown under minimal assumptions, but it is not tight. To estimate the true generic decay rate, we turn to hydrodynamic reasoning. One expects a height-field minimum to host coherence until it encounters a deeper minimum through transport. For windows of size $\ell$, the deepest neighboring minimum is $\sim \ell$ away, so the window minimum will persist for $t_\ell \sim \ell^z$ with $z$ the spin transport dynamical exponent. While the minimum persists, it can wander, and $C(t)$ is controlled by its return probability $\sim t^{-1/z}$. Ignoring other contributions (which we expect to decay exponentially), we expect
\begin{equation}
C(t) \simeq \rho w(t) P_{\mathrm{ret}}(t) \sim t^{-2/z},
\label{eq:structure}
\end{equation}
where $\rho$ is the probability that $\sigma^+$ attaches to a height minimum, $w(t)$ is the survival probability of that minimum, and $P_{\mathrm{ret}}(t)$ is the probability that the minimum returns to its original position at time $t$. In chaotic systems (namely, the $\beta \neq 0$ models), both the lifetime of a minimum and its return probability are set by the transport exponent $z$, which is $\approx 8/3$ for the Fredkin chain~\cite{SWVF2021,MSGV2025} and generically $2$. The expectation that the transverse autocorrelator decays as the \emph{square} of the standard density autocorrelator is borne out by numerics even on timescales when the latter has not reached its asymptotic value [Fig.~\ref{fig:mazur}(c)]. As we discuss next, however, exact or approximate integrability can cause both coefficients to decay slower than this simple estimate predicts.

\begin{figure}[t]
{
\centering
\includegraphics[width=\columnwidth]{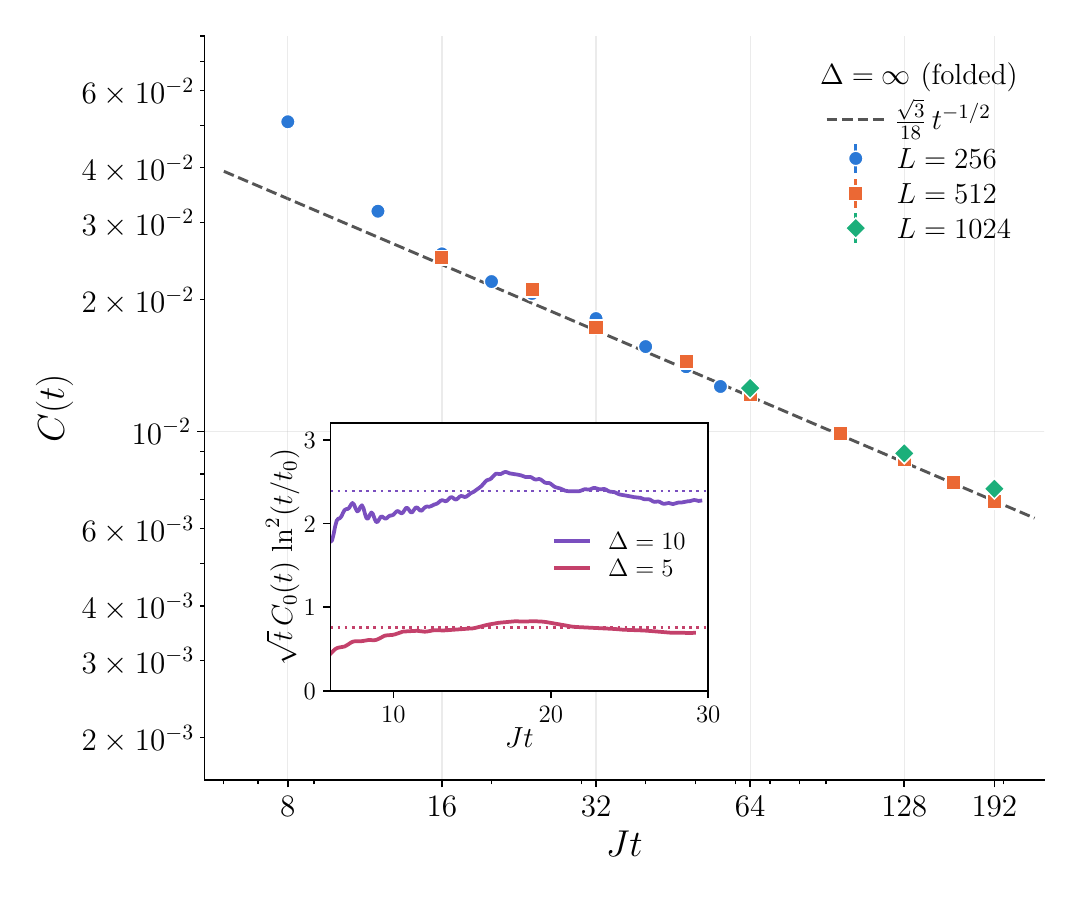}
\caption{{\bf Transverse correlator in the XXZ chain.} The folded ($\Delta=\infty$) chain correlator, computed by unbiased Monte Carlo sampling of free-fermion Slater determinants using the results of Refs.~\cite{Pozsgay,Hough2006,Sun2023} for open chains with $L=256$, $512$ and $1024$ (up to $10^6$ samples), compared with $\frac{\sqrt3}{18}\,t^{-1/2}$. Inset: XXZ chains with $\Delta=10$ and $5$, computed by dynamical typicality~\cite{Bartsch2009,Heitmann2020} on periodic chains with $L=24$. $C_0(t)$ is the slow part of $C(t)$, obtained by fitting $C(t)=C_0(t)+C_1(t)\cos\Delta t+S_1(t)\sin\Delta t$ in a sliding window, and $t_0$ comes from the fit $\sqrt{t}\,C_0=a/\ln^2(t/t_0)$ over $12\le Jt\le 29$, giving $t_0=0.11$ and $a=2.39$ for $\Delta=10$ and $t_0=0.76$ and $a=0.76$ for $\Delta=5$ (dotted lines show $a$).}
\label{fig:folded}
}
\end{figure}

\phead{Folded XXZ chain}
We begin with the folded ($\Delta = \infty$) limit of the XXZ spin chain, setting $J=1$.
In this model, $\sg^+$ can act on a configuration in one of two ways. First, it can create a magnon (e.g., $\downarrow\downarrow\downarrow \implies \downarrow \uparrow \downarrow$), which will propagate through the system and cause dephasing when the other magnons ``measure'' it through collisions. These contributions to $C(t)$ decay exponentially and are unimportant at late times. Alternatively, $\sg^+$ can modify the frozen pattern (e.g., $\downarrow \downarrow \downarrow\uparrow \uparrow \implies \downarrow \downarrow \uparrow\uparrow \uparrow$). Magnons move identically through every frozen pattern: thus, a change in the frozen pattern affects $C(t)$ only insofar as the pattern itself has shifted. In terms of Eq.~\eqref{eq:structure}, therefore, $w(t) = 1$. The return probability of the domain wall is $P_{\mathrm{ret}}(t) \sim 2/\sqrt{4\pi D t}$, where $D = 4/(3\pi)$ is the infinite-temperature spin diffusion constant of the folded XXZ chain~\cite{GV, SciPostPhys.6.4.049}, and the factor of two comes from a lattice even-odd effect: the frozen pattern only moves by two steps, so it always lives on the same sublattice, and its return probability is thus twice the hydrodynamic estimate.

Finally, the probability of exciting the frozen pattern without creating magnons is $\rho = 1/9$, a combinatorial factor we obtain as follows. For $\sg^+_x$ to act without changing the number of magnons, the following conditions must hold. The spin at site $x$ must be $\downarrow$ in the first place, and its neighbors must be anti-aligned (since otherwise a magnon would be either created or destroyed). These conditions hold with probability $1/4$. Further, flipping $x$ should not create or destroy a magnon on either side of it. For example, if the spin configuration around the central site $x = 0$ is $\uparrow_{-1} \downarrow_{0} \downarrow_{1} \uparrow_{2} \uparrow_{3}$, flipping $x = 0$ gives $\uparrow_{-1} \uparrow_{0} \downarrow_{1} \uparrow_{2} \uparrow_{3}$, creating a magnon immediately to its right.
Configurations in which a magnon is created or destroyed to the right can be identified by the following rule~\cite{SM}: move right from $x+1$ to the nearest pair of aligned spins. Count the intervening anti-aligned pairs. The magnon number changes iff this number is odd (as in the example above). Conditional on the local requirements above, the probability that the spin-flip does not change the magnon number to its right is thus $2/3$. Combining this with the analogous requirement to the left, the total probability of not creating or destroying a magnon is $\rho = (1/4)(2/3)^2 = 1/9$. Putting everything together,
\begin{equation}
C(t) = \frac{1}{9}\cdot\frac{\sqrt3}{2}\,t^{-1/2} + \ldots \approx0.096\,t^{-1/2},
\label{eq:Ctotal}
\end{equation}
where $\ldots$ denotes faster-decaying terms. This prediction is in excellent agreement with numerics on chains of up to $1024$ sites~(Fig.~\ref{fig:folded}).

\phead{XXZ for $1<\Delta<\infty$}
Away from $\Delta=\infty$, there is no strictly frozen pattern. Instead, the XXZ model supports a hierarchy of stable magnon bound states, or ``strings,'' described by the thermodynamic Bethe ansatz~\cite{TakahashiBook}. A bare $s$-string is a bound state of $s$ overturned spins in a polarized background: at large anisotropy, it resembles a domain of $s$ aligned spins. At infinite temperature and zero magnetization, the density of these strings scales as $\rho_s\sim s^{-3}$ at large $s$, while their characteristic velocities are exponentially suppressed, $v_s\sim e^{-\eta s}$, where $\cosh\eta=\Delta$ and $\eta=\ln q$~\cite{Bertini2016,CastroAlvaredo2016,gopalakrishnan2019anomalous}. Generalized hydrodynamics~\cite{Bertini2016,CastroAlvaredo2016} describes their ballistic propagation, together with diffusive broadening induced by fluctuations in collisions with other strings~\cite{PhysRevB.98.220303,PhysRevLett.121.160603,SciPostPhys.6.4.049}. For heavy strings, the diffusion constant approaches the spin diffusion constant $D(\Delta)$~\cite{GV,gopalakrishnan2019anomalous}. At large anisotropy, this diffusion is dominated by collisions with magnons. Although intrinsic string motion becomes appreciable on a hopping timescale of order $e^{\eta s}$, ballistic displacement dominates diffusive broadening only on the parametrically longer timescale $D(\Delta)/v_s^2\sim e^{2\eta s}$.

We focus on the long-time contribution in which $\sg^-$ changes the length of a heavy string by one. The corresponding weight is proportional to $\rho_s$ at large $s$, $a_s\sim2/s^3$, as obtained from explicit enumeration~\cite{SM}. Before intrinsic hopping becomes appreciable, the two string types respond approximately identically to collisions with lighter strings, preserving their relative coherence. Once they undergo different intrinsic motion, their scattering histories differ and this contribution dephases. We therefore estimate a coherence time $\tau_s\sim t_0e^{\eta s}$, where $t_0$ is a microscopic timescale. Approximating the coherence factor $w_s(t)$ by a cutoff at $t\sim\tau_s$, the surviving strings have lengths $s>s^*(t)\sim\ln(t/t_0)/\eta$. Inserting their weights into Eq.~\eqref{eq:structure} gives
\begin{equation}
C(t)\simeq P_{\mathrm{ret}}(t)\sum_s a_s w_s(t)
\sim\frac{1}{\sqrt{D(\Delta)t}}
\left[\frac{\ln q}{\ln(t/t_0)}\right]^2.
\label{finite_Delta}
\end{equation}
This case differs qualitatively from $\Delta = \infty$: here $C(t)$ decays parametrically faster than $\langle \sigma^z_x(t) \sigma^z_x(0)\rangle$, which is strictly diffusive. The motion of a heavy string leads to backflow for yet heavier strings; this backflow contributes to hydrodynamics, but does not affect dephasing. This prediction is consistent with numerics, although the logarithmic factor cannot be isolated at the times we can reach~(Fig.~\ref{fig:folded}).

\phead{Away from half-filling}
So far we have considered ensembles that are invariant under global spin-flip. We briefly consider infinite-temperature (product-state) ensembles that have a net magnetization density $\langle \sigma^z_x \rangle = m$. This modification immediately makes the Mazur bound~\eqref{eq:lcbound} trivial for any $m \neq 0$, for a reason that is clear in the height representation: net magnetization corresponds to a net tilt in the height representation, and when $\ell \gg 1/m^2$, the height minimum of a region of size $\ell$ is overwhelmingly likely to occur at the region's edge. The operator $\et$ is localized at one edge of the window, so locality does not protect it for longer than an $O(1)$ timescale. In the chaotic chains, therefore, one expects a crossover to stretched-exponential dephasing~\cite{MJKG2025, MJG2026} on length-scales $\ell \agt m^{-2}$, or equivalently timescales $t \agt m^{-2z}$. The situation is particularly clear for the Fredkin chain, where $m$ of order unity allows one to neglect the Fredkin constraint entirely (since there are no deep local valleys left), and the dynamics should be completely generic. This sensitivity to $m$ illustrates that the quantum group symmetry is sharply distinct from, say, the conventional continuous $SU(2)$ symmetry that obtains at the Heisenberg point: the latter guarantees hydrodynamics for conserved densities regardless of the background ensemble. 

This crossover to (stretched-)exponential dephasing is absent in the integrable quantum group models. In the extreme case of the folded XXZ spin chain, nothing changes relative to half filling except the densities of magnons and frozen patterns. The transverse correlator still decays diffusively, with the exact expression $C(t) =  \frac{1-m}{2}\Big[\frac{1-m^2}{3+m^2}\Big]^{3/2}t^{-1/2}$.
At finite $\Delta$, nonzero magnetization exponentially suppresses
large strings: $\rho_s\sim e^{-2|\mu|s}$, where
$\mu=\operatorname{arctanh}m$. Using the coherence cutoff
$s^*(t)\sim\ln(t/t_0)/\ln q$, our kinetic argument therefore predicts,
up to possible logarithmic factors,
$C(t)\sim t^{-1/2-2|\mu|/\ln q}$.
\phead{Discussion}
Slow dephasing of transverse correlators is unexpected even in integrable models: for example, at the $\Delta = 0$ limit of the XXZ model (which maps to free fermions), $C(t)$ decays superexponentially at infinite temperature~\cite{BJ,CapelPerk,GKSS}. 
That it occurs in chaotic models with diffusive hydrodynamics is therefore doubly surprising: at modest values of $q, \beta$, nothing else about the dynamics generated by $H(\beta)$ is slow or anomalous: $C(t)$ decays slowly because the operator $\R$  locks its dynamics to a hydrodynamic mode. 
In the models we explored, this locking ultimately came purely from a quantum group symmetry rather than from integrability (unlike the quasilocal charges of Refs.~\cite{Prosen2011,ProsenIlievski2013,PPSA2014,IMPZ2016}). Whether it occurs more generally for constrained or frustration-free Hamiltonians---especially those in higher dimensions---is an interesting question for future work. 
Our results raise other natural questions as well: whether the ``non-undershoot'' can be proved for some class of ``natural'' many-body dynamics (we are not aware of any systems that violate it); whether experimentally tractable $U_q(sl_2)$-symmetric quantum circuits can be devised, and used to search for persistent quantum coherence; and whether the quantum-group perspective can shed light on still unsolved questions about the dynamics of the Fredkin spin chain and other constrained spin chains. 

\begin{acknowledgments}
We thank Luca Delacr\'etaz, Jacopo De Nardis, Viktor Gorbenko and Aleksandr Zhabin  for helpful discussions. This work involved extensive discussions with Claude Fable 5.1, ChatGPT Astra 6 and Sol 6.1, which also coded up the numerical tests. S.G. was supported by the NSF Institute for Robust Quantum Simulation. R.V. was supported by the Swiss National Science Foundation (grant 10008234) and by the Foundation for the University of Geneva. 
\end{acknowledgments}

\bibliography{refs}

\bigskip
\onecolumngrid
\newpage


\section*{Supplementary material}

# Supplemental Material

## Long-lived local coherences at high temperature from frozen domain walls

Sarang Gopalakrishnan[1] and Romain Vasseur[2,3]
[1] *Department of Electrical and Computer Engineering, Princeton University, Princeton, NJ 08544, USA*
[2] *Department of Theoretical Physics, University of Geneva, 24 quai Ernest-Ansermet, 1211 Genève, Switzerland*
[3] *Geneva Quantum Center, University of Geneva*

### CONTENTS



### I. QUANTUM-GROUP SYMMETRY AND TEMPERLEY–LIEB CHAINS

In this section we summarize the quantum-group representation and the local operators that commute with it. These give the conserved polar charge used in the overlap bound and ensure that the same construction applies on a finite window.

We use the notation of the main text: $S_j^z = \sigma_j^z/2$, $\sigma_j^{\pm} = (\sigma_j^x \pm i\sigma_j^y)/2$, and $\langle O \rangle = 2^{-L}\,\mathrm{Tr}\,O$ at infinite temperature. Sites are numbered $1, \ldots, L$. For a configuration $c$, $h_c(i) = \sum_{j<i} s_j(c)$ is the height before site $i$, and $d(c)$ is the down step that first reaches the global minimum; $d(c) = 0$ if the minimum is at the left boundary.

The quantum group $U_q(sl_2)$ is generated by $E$, $E^\dagger$ and $K = q^{S^z}$, with

$$KEK^{-1} = qE, \qquad KE^\dagger K^{-1} = q^{-1}E^\dagger, \qquad [E, E^\dagger] = \frac{K^2 - K^{-2}}{q - q^{-1}}. \tag{1}$$

For one spin, $E = \sigma^+$ and $K = q^{\sigma^z/2}$. The deformation changes the rule for combining spins: on $V_a \otimes V_b$, where $V_a$ and $V_b$ are spin-$a$ and spin-$b$ multiplets,

$$E_{a,b} = E_a \otimes K_b + K_a^{-1} \otimes E_b, \qquad K_{a,b} = K_a \otimes K_b. \tag{2}$$

This rule for combining representations is called the coproduct. Iterating it gives the charge in Eq. (2) of the main text,

$$E = \sum_{i=1}^{L} q^{-\sum_{k<i} S_k^z} \sigma_i^+ q^{\sum_{k>i} S_k^z}. \tag{3}$$

The $q$-numbers are $[n]_q = (q^n - q^{-n})/(q - q^{-1})$, with $[n]_1 = n$. For real $q > 0$, the chain decomposes into spin-$j$ representations with the same multiplicities as for $SU(2)$ [1]. In an orthonormal basis of $V_j$,

$$K_j|j, m\rangle = q^m|j, m\rangle, \qquad E_j|j, m\rangle = \lambda_j(m)|j, m+1\rangle, \qquad \lambda_j(m) = \sqrt{[j-m]_q[j+m+1]_q}. \tag{4}$$

The polar charge $\mathcal{R}$ replaces the coefficient $\lambda_j(m)$ by one on each allowed transition and annihilates highest-weight states. More generally, we use the canonical polar factor

$$\mathrm{Pol}(A) = Af(A^\dagger A), \qquad f(s) = \begin{cases} s^{-1/2}, & s > 0, \\ 0, & s = 0. \end{cases} \tag{5}$$

Thus $\mathcal{R} = \mathrm{Pol}(E)$. In the limit $q \to \infty$, its action becomes particularly simple in the spin-product basis. Write an up spin as an opening bracket and a down spin as a closing bracket, and match brackets in the usual way, allowing nested pairs. The unmatched spins then consist of down spins followed by up spins. The limiting operator $\tilde{e}$ flips the last unmatched down spin and annihilates a configuration if there is no unmatched down spin [2, 3]. In the height representation, this is precisely the down step that first reaches the global minimum. This limiting, unit-amplitude action is what is meant here by the *crystal raising operator*; below we generally refer to it simply as the limiting raising operator. If $H$ is Hermitian and $[H, E] = 0$, then $H$ commutes with $E^\dagger E$ and hence with $\mathcal{R}$.

The local operators that commute with this quantum-group action are generated by the Temperley–Lieb operators

$$e_j = (q + q^{-1})|\psi_q\rangle\langle\psi_q|_{j,j+1}, \qquad |\psi_q\rangle = \frac{q^{1/2}|\uparrow\downarrow\rangle - q^{-1/2}|\downarrow\uparrow\rangle}{\sqrt{q + q^{-1}}}. \tag{6}$$

They obey $e_j^2 = (q + q^{-1})e_j$, $e_j e_{j\pm1} e_j = e_j$, and $[e_i, e_j] = 0$ for $|i - j| > 1$. In Pauli notation,

$$e_j = -\frac{1}{2}(\sigma_j^x\sigma_{j+1}^x + \sigma_j^y\sigma_{j+1}^y) - \frac{\Delta}{2}\sigma_j^z\sigma_{j+1}^z + \frac{q - q^{-1}}{4}(\sigma_j^z - \sigma_{j+1}^z) + \frac{\Delta}{2}, \qquad \Delta = \frac{q + q^{-1}}{2}. \tag{7}$$

The single-spin terms telescope in a sum over bonds. Consequently, $-2\sum_j e_j$ is the XXZ chain with the boundary fields required by quantum-group symmetry, up to a constant. Any Hermitian polynomial in the $e_j$ commutes with $E$, including the three-site family in Eq. (1) of the main text. This remains true on any open interval, with its own coproduct charge. Removing all terms that cross the ends of an interval therefore gives a symmetric window Hamiltonian.

For comparison with Ref. [4], the numerical data below use

$$H_{\mathrm{QG}} = -2\sum_{j=1}^{L-1} e_j + \lambda\sum_{j=1}^{L-2} V_j, \qquad V_j = 2(e_j e_{j+1} + e_{j+1} e_j) - 2\Delta(e_j + e_{j+1}), \tag{8}$$

with irrelevant constants omitted. In the bulk this is $(1 + 2\lambda\Delta)H(\beta)$, with

$$\beta = \frac{2\lambda}{1 + 2\lambda\Delta}. \tag{9}$$

For an open chain the two expressions differ additionally by the symmetric edge term $-(e_1 + e_{L-1})$. The XXZ flip-flop amplitude in Eq. 8 is 2; the folded-chain discussion below instead keeps the hopping scale $J$ explicit.

## II. STATIC MAZUR OVERLAP AT FINITE $q$

The Mazur bound in Eq. (4) of the main text requires the static overlap $\langle\sigma_x^-\mathcal{R}\rangle$ to decay algebraically, not exponentially, with $L$. At $q=\infty$ this overlap is a random-walk probability: $\mathcal{R}=\tilde{e}$ flips exactly one spin, the down step that first reaches the height minimum, so

$$\langle\sigma_x^-\tilde{e}\rangle = p_{L,x} = 2^{-L}\binom{x-1}{\lfloor (x-1)/2\rfloor}\binom{L-x}{\lfloor (L-x)/2\rfloor}, \tag{10}$$

which counts configurations whose height profile stays weakly above its value at $x$ on both sides (before its first minimum the reversed walk is weakly nonnegative, and so is the walk after it). For $x/L\to\alpha\in(0,1)$, $p_{L,x}\sim[\pi L\sqrt{\alpha(1-\alpha)}]^{-1}$, which is $2/(\pi L)$ at the center. In this section we explain intuitively why finite $q$ does not spoil this algebraic decay, compute the leading finite-$q$ correction, and verify the resulting bound numerically.

### A. Intuitive argument

A useful physical picture for the persistence of the $1/L$ overlap at finite $q$ comes from the height dependence of the raising generator. For a product configuration $c$ with a down spin at $x$ and total magnetization $M$,

$$\langle c|\sigma_x^- E|c\rangle = q^{M+1/2-h_{\min}}\, q^{-(h_c(x)-h_{\min})}. \tag{11}$$

The first factor varies exponentially between configurations and suppresses the normalized overlap of $E$ [4], whereas the second favors spins near the global height minimum. The polar decomposition removes the varying ladder amplitudes within each quantum-group multiplet. If $\mathcal{R}$ retains this concentration near the minimum, its effective support is restricted to a height range of order $1/\ln q$. At $q=\infty$, $\mathcal{R}$ flips only the down step that first reaches the minimum. At finite $q$, this single-site action is broadened over a finite effective number of sites. Random-walk scaling suggests a number of sites of order $1/\ln^2 q$ when this scale exceeds the lattice spacing, but independent of $L$ at fixed $q>1$. Since a specified bulk site lies near the global minimum with probability of order $1/L$, this picture suggests that $\langle\sigma_x^-\mathcal{R}\rangle$ retains its $1/L$ scaling, with a $q$-dependent prefactor. It also explains the crossover near $q=1$: when $1/\ln q$ becomes comparable to the typical height excursion $\sqrt{L}$, the raising action spreads throughout the chain, consistent with a $L^{-1/2}$ overlap at the $SU(2)$ point. The expansion below gives a direct check at large $q$ and shows that the leading correction enhances the overlap.

### B. Large $q$ expansion

There is a simple reason why moving away from $q=\infty$ can only increase the overlap at leading order. In the quantum-group multiplet basis, $\mathcal{R}$ raises the magnetization with unit amplitude. The overlap $\mathrm{Tr}(\sigma_x^-\mathcal{R}) = \sum_{c:s_x=\downarrow}\langle c|\sigma_x^-\mathcal{R}|c\rangle$ is a sum of spin-flip amplitudes in the product basis. At $q=\infty$ the amplitude is 1 if $x=d(c)$ is the first-minimum step and 0 otherwise. As the expansion below shows, the first-minimum flip receives no correction at order $1/q$, while additional single-spin flips acquire positive amplitudes at this order. Since the overlap sums amplitudes rather than their squares, the leading correction is a positive count of configurations. This is the opposite of what happens for the bare generator $E$, whose matrix elements carry the exponentially varying weights $q^{-h_c(x)}$ of Eq. (2) of the main text. Those weights, not the structure of the multiplets, make the normalized overlap of $E$ exponentially small [4].

To make this quantitative, expand in $q^{-1}$ at fixed $L$. At large $q$ the quantum-group multiplets are explicit in the product basis. Match up and down spins as brackets (Sec. I). Each matched pair $(a,b)$, with $a<b$, is in the $q$-singlet $|\uparrow_a\downarrow_b\rangle - q^{-1}|\downarrow_a\uparrow_b\rangle$. The unmatched spins, which always consist of some down spins followed by some up spins, form a single spin-$j$ multiplet with $2j$ equal to their number: each of its states is, to leading order, a product configuration $\downarrow\cdots\downarrow\uparrow\cdots\uparrow$ on the unmatched sites, and its first correction, of relative weight $+q^{-1}$, exchanges the last of the unmatched down spins with the first of the unmatched up spins. (This is the large-$q$ form of the $q$-symmetric states: applying $E$ to $\downarrow\cdots\downarrow$ flips the last site with the largest weight.) Writing a product state $|c\rangle$ in terms of these multiplet states therefore involves a change of basis $V=1+q^{-1}B+O(q^{-2})$ with

$$B|c\rangle = -\sum_{\text{matched pairs }(a,b)} |c^{ab}\rangle + |c^{du}\rangle, \tag{12}$$

where $c^{ab}$ denotes $c$ with the spins at $a$ and $b$ exchanged, $d$ and $u$ are the last unmatched down spin and the first unmatched up spin of $c$, and the second term is absent if $c$ has no unmatched down spin or no unmatched up spin. In the multiplet basis $\mathcal{R}$ is exactly the unit ladder $\tilde{e}$, so

$$\mathcal{R} = V\tilde{e}V^{-1} = \tilde{e} + q^{-1}[B,\tilde{e}] + O(q^{-2}). \tag{13}$$

(The basis defined by $V$ is not orthogonal, but its Gram matrix acts only on the multiplicity labels of the multiplets and commutes with $\tilde{e}$, so the polar factor is insensitive to it.)

The matrix element $\langle \sigma_x^+ c|[B,\tilde{e}]|c\rangle$ is nonzero, and then equal to $+1$, in exactly two situations. Either $x$ is the *second-to-last* unmatched down spin of $c$, the step that first reaches one unit above the global minimum, and the spin-$j$ admixture moves the raising action from the last unmatched down spin onto it; or $x$ is the down spin of a matched pair $(a,x)$ that closes at the minimum, and the singlet admixture exchanges the pair, making $a$ the last unmatched down spin, so that raising $a$ in the exchanged configuration yields $\sigma_x^+ c$. In the bulk both are as likely as the first-minimum flip itself. In the second-to-last-step event, the walk to the left of $x$ stays weakly above its value at $x$ and the walk to the right must dip exactly one unit below it (the global minimum is still to come). In the closing-pair event, the walk to the left must reach one unit below its value at $x$ without going lower (the minimum has already been visited), and the walk to the right may either stay weakly above the endpoint or dip exactly one unit below it. Each of these constraints has, asymptotically, the probability of a nonnegative walk, except the last "either/or", which has twice that probability; constraints that differ only by excluding bridges are equivalent up to $O(1/L)$. The first event therefore contributes once and the second twice, so

$$\langle \sigma_x^- \mathcal{R}\rangle = p_{L,x}\left[1 + \frac{3}{q} + O(q^{-2})\right] \qquad (x,\, L-x \to \infty). \tag{14}$$

With boundaries the same reflection-principle count gives the coefficient $c_{L,x} = 3 - 2r_n - 2r_m + r_n r_m$ in place of 3, with $n = x-1$, $m = L-x$, $r_k = S_k/Q_k$, $Q_k = \binom{k}{\lfloor k/2\rfloor}$ and $S_k = Q_k/(k/2+1)$ for even $k$ (zero for odd $k$); it is strictly positive for $L \geq 2$ and bounded by three. We have checked it against the exact overlap at $q = 200$ for all sites of chains with $L \leq 7$; the coefficients agree to three digits. The expansion is at fixed $L$ and does not by itself control the remainder uniformly in $L$; that is what the numerics below establish.

### C. Numerical verification and the Mazur bound

We have computed the overlap exactly, sector by sector, for all sites $x$ of open chains with $L \leq 14$ and various values of $q$. In every case it lies between its two exactly known endpoints, $p_{L,x} \leq \langle \sigma_x^- \mathcal{R}\rangle \leq \langle \sigma_x^- \mathcal{R}\rangle|_{q=1}$, where at $q = 1$ the polar factor $\mathcal{R}$ acts as the unit ladder within each $SU(2)$ multiplet, so that $\langle \sigma_x^- \mathcal{R}\rangle\big|_{q=1} = \frac{1}{L\,2^L}\sum_j m_j^{(L)} \sum_{m=-j}^{j-1}\sqrt{(j-m)(j+m+1)} \simeq \sqrt{\frac{\pi}{8L}}$, with $m_j^{(L)}$ the multiplicities of Eq. (A29). The overlap decreases monotonically with $q$ at every site, and it approaches $p_{L,x}$ with the slope predicted by Eq. (14). The rank of $\mathcal{R}$ reproduces Eq. (16) below in every case. Our exact diagonalization data are consistent with this picture. In particular

$$\langle \sigma_x^- \mathcal{R}\rangle \geq p_{L,x} \qquad (q \geq 1) \tag{15}$$

holds in every case tested, and we are not aware of any mechanism by which the overlap could become exponentially small at intermediate $q$. The inequality (15) can in fact be proved for all $L$ and all $q \geq 1$. The proof represents the polar factors of the raising operators on quantum-group multiplets as zero-temperature magnetizations of transverse-field Ising chains and uses Griffiths-type correlation inequalities to compare them. It is reproduced in Appendix A. We note that this argument was generated with the assistance of large language models with minimal human input, and has been formalized and machine-checked in Lean 4 (Mathlib). We have instead verified each of its intermediate inequalities numerically and its conclusion is corroborated by the exact results above. Since the physics of the bound is contained in the elementary discussion of this section, we have placed the proof in an appendix for the interested reader.

The Mazur–Suzuki inequality [5, 6], Eq. (3) of the main text, applies to the conserved charge $\mathcal{R}$. There is one highest-weight state per multiplet, giving the $q$-independent denominator

$$\langle \mathcal{R}^\dagger \mathcal{R}\rangle = 1 - 2^{-L}\binom{L}{\lfloor L/2\rfloor}. \tag{16}$$

Eq. (15) therefore gives, for all $q \geq 1$

$$\overline{C} \geq \frac{|\langle \sigma_x^- \mathcal{R}\rangle|^2}{\langle \mathcal{R}^\dagger \mathcal{R}\rangle} \geq \frac{p_{L,x}^2}{1 - 2^{-L}\binom{L}{\lfloor L/2\rfloor}}, \tag{17}$$

which at the center reproduces the $4/(\pi^2 L^2)$ bound in Eq. (4) of the main text. The infinite-time correlator remains above this bound in both the integrable XXZ chain and the chaotic quantum-group and Fredkin chains. At fixed $L$ and $q$, the static bound is independent of the Hamiltonian within the quantum-group-symmetric family, whereas the plateau depends on the dynamics. The calculation of the overlaps and plateaus is described in Sec. VI.

## III. NON-UNDERSHOOT ASSUMPTION AND THE FINITE-TIME BOUND

In this section we explain how the finite-size Mazur bound constrains the late-time correlator in an infinite chain. The locality estimate relates the infinite chain to a finite window, but the Mazur bound controls only the infinite-time average within that window. An additional assumption about the approach to the plateau is therefore needed. We state this assumption explicitly, derive its consequence for the interval-averaged correlator, and describe the numerical checks. Let $W = [x-\ell, x+\ell]$, and retain only Hamiltonian terms supported entirely within $W$. By Sec. I, $H_W$ commutes with the window charge. Equation 17 therefore gives

$$\overline{C}_W \geq B_\ell \equiv \frac{p^2_{2\ell+1,\ell+1}}{1-2^{-(2\ell+1)}\binom{2\ell+1}{\ell}} \sim \frac{1}{\pi^2\ell^2}. \tag{18}$$

### A. Locality and the Averaging Interval

The terms omitted from $H_W$ either lie outside the window or cross its ends. Only the latter contribute to the difference

$$\sigma^-_x(t)|_H - \sigma^-_x(t)|_{H_W} = i\int_0^t ds\, e^{iHs}[H-H_W, e^{iH_W(t-s)}\sigma^-_x e^{-iH_W(t-s)}]e^{-iHs}. \tag{19}$$

The Lieb–Robinson bound [7–9] then gives

$$|C(t) - C_W(t)| \leq K J_* t\, e^{(vt-\ell)/\xi} \equiv \epsilon(\ell, t), \tag{20}$$

where $J_*$ bounds the norm of a local term and $K, \xi$ are independent of $\ell, t$. For Eq. 8, $J_* = O(q + |\lambda|q^2)$. The large diagonal terms do not set the propagation velocity: in their interaction picture, each off-diagonal term acquires local diagonal phases but retains its norm. Its support expands by only a fixed number of sites. Applying the Lieb–Robinson bound to this time-dependent Hamiltonian gives $v = O(1 + |\lambda|q)$. For the Fredkin chain, $v = O(1)$.

We average over the interval $[t, 2t]$,

$$\tilde{C}_W(t) = \frac{1}{t}\int_t^{2t} C_W(\tau)\, d\tau. \tag{21}$$

In the following inequalities $C$ denotes its real part. If $\operatorname{Re} C_W(t) = \overline{C}_W + \sum_{\omega\neq 0} S_W(\omega)\cos\omega t$, the oscillatory contribution to this average is

$$\tilde{C}_W(t) - \overline{C}_W = \sum_{\omega\neq 0} S_W(\omega)\frac{\sin 2\omega t - \sin\omega t}{\omega t}. \tag{22}$$

This suppresses rapid oscillations, but its kernel changes sign. Thus $\tilde{C}_W(t) \geq \overline{C}_W$ does not follow from the positivity of the spectral weights. Positive kernels centered at $t = 0$ avoid this problem, but include the short-time transient and do not give the late-time bound required here.

Choose $\ell(t) = 2vt + O(\ln t)$ so that $\epsilon(\ell, 2t) = o(t^{-2})$. The assumption is that the smoothed window correlator has not fallen below its plateau on this time scale:

$$\tilde{C}_{W(\ell(t))}(t) \geq \overline{C}_{W(\ell(t))}. \tag{23}$$

Averaging Eq. 20 over $[t, 2t]$ and using Eq. 18 gives

$$\tilde{C}(t) \geq \tilde{C}_W(t) - \epsilon(\ell, 2t) \geq B_\ell - \epsilon(\ell, 2t) = \frac{1+o(1)}{4\pi^2 v^2 t^2}. \tag{24}$$

A weaker assumption $\tilde{C}_W \geq \kappa\overline{C}_W$, with fixed $\kappa > 0$, gives the same power with a factor $\kappa$. Either form excludes a decay faster than $t^{-2}$ throughout every sufficiently late averaging interval.

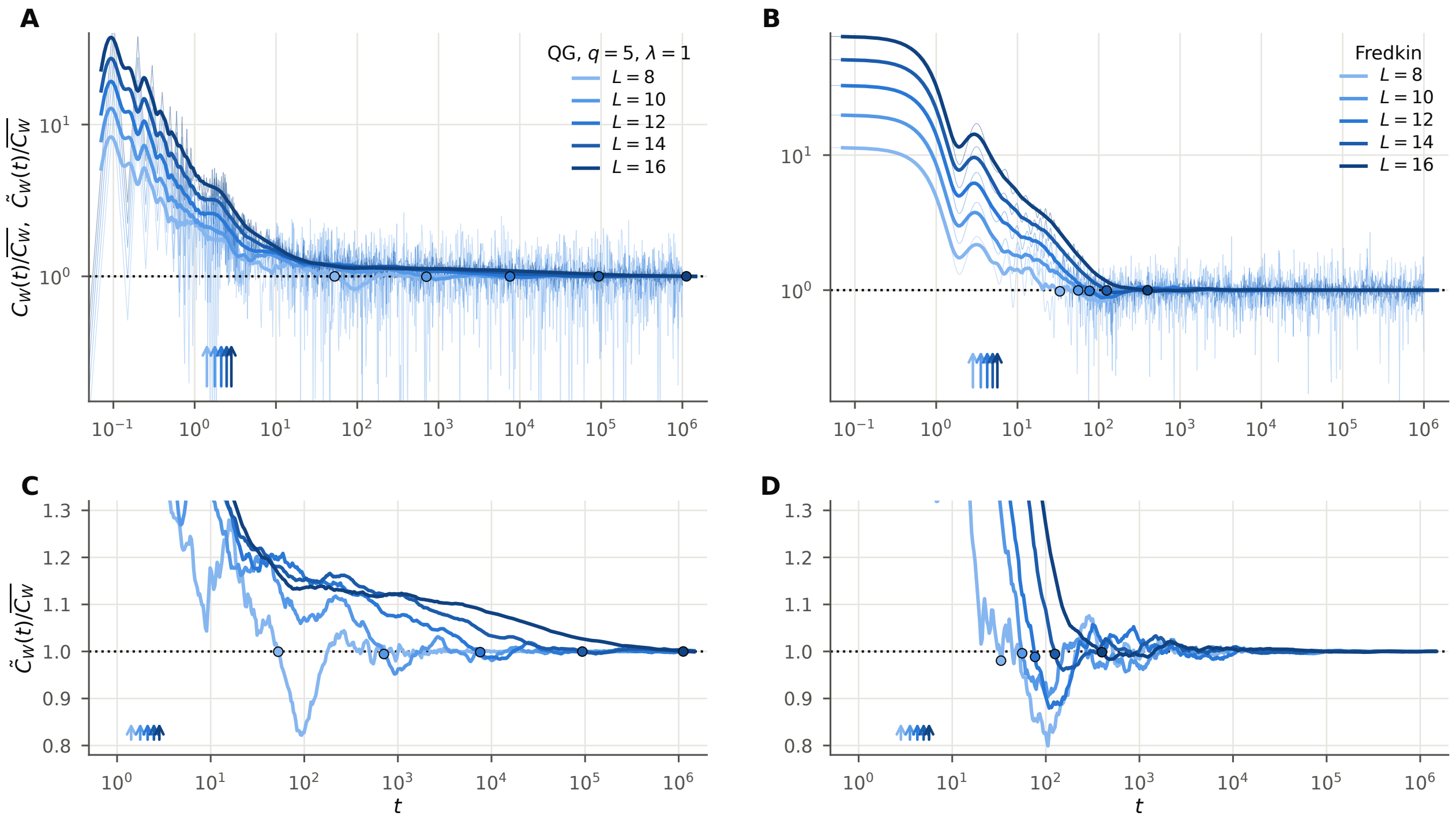


FIG. 1. **Check of the non-undershoot assumption.** (A) Quantum-group chain at $q = 5$, $\lambda = 1$: $C_W(t)$ (thin), $\tilde{C}_W(t)$ (thick) and $\overline{C}_W$ (dotted), normalized by $\overline{C}_W$. The interval average is plotted at $\sqrt{2}\,t$. (B) The same for the Fredkin chain. (C,D) Interval averages on a linear scale. Arrows mark $t = \ell/(2v)$, with $\ell = L/2$ and $v = 2$ in (A), $v = 1$ in (B); a larger velocity moves these arrows to earlier times. Circles mark the first interval with $\tilde{C}_W(t) < \overline{C}_W$.

### B. Physical Motivation and Numerical Checks

In the height picture, the minimum of the window initially explores only a small fraction of it. If its displacement scales as $t^{1/z}$, the correlator associated with that minimum is of order $\ell^{-1}t^{-1/z}$, while its plateau is of order $\ell^{-2}$. At the light-cone time $t \sim \ell/v$, their ratio is $\ell^{1-1/z} \gg 1$ for $z > 1$. This motivates Eq. (23): the minimum has not equilibrated over the window when locality is used. It is a kinetic estimate, not a proof of the assumption.

Figure 1 compares $C_W(t)$, its interval average, and its infinite-time average on open chains of $L = 8$–$16$. The interval average eventually crosses below the plateau, but only at times beyond those needed for the bound. A check independent of a chosen velocity uses the first time $t_{\rm lc}$ at which the $L$- and $(L+2)$-site correlators differ by $\overline{C}_W/2$. At $q = 5$, $\lambda = 1$, $\tilde{C}_W(t_{\rm lc}/2)/\overline{C}_W$ ranges from 3.2 to 5.4 for $L = 8$–$14$. The corresponding ratios range from 1.1 to 4.8 in the Fredkin chain and in the other quantum-group chains tested ($q = 3, 10$, and $\lambda = 0$). The late-time pointwise oscillations are substantially larger than those of the interval average; at $q = 5$ their standard deviation relative to the plateau falls from 0.53 at $L = 8$ to 0.06 at $L = 16$.

## IV. XXZ CHAIN

In this section we give the counting underlying the long-time transverse correlator in the XXZ chain. We first discuss the folded limit, in which the spin flip modifies a frozen pattern, and then extend this picture to the hierarchy of strings at finite anisotropy.

### A. Folded Limit

At $\Delta = \infty$, only spin exchanges that preserve the Ising energy remain in the slow dynamics: neighboring opposite spins can exchange if their two outer neighbors are equal. This constrained Hamiltonian is the folded XXZ chain. Its configurations separate into mobile lone spins and a residual sequence of larger domains. In the representation

used here, each lone spin is grouped with a neighboring partner into a two-site object, which we call a rod. Once the excluded length of each rod is removed, their motion is equivalent to that of free fermions on a shorter chain [10]; we call its site coordinate $y$ and retain the hopping amplitude $J/2$. The residual spin sequence cannot change through this motion, but a rod passing a domain wall shifts that wall by two physical sites. Thus "frozen" refers to the order and identity of the residual domains, rather than to their positions, and a marked wall moves only by even displacements.

A spin flip contributes to the long-lived channel when it shifts such a marker without creating or absorbing a rod. The parity rule and its Bernoulli counting are given in Sec. IV D. At half filling the total weight is $\rho(2) = 1/9$. The pattern diffusion constant is $D = 4J/(3\pi)$ [11–13], and the even-displacement return probability is $2/\sqrt{4\pi Dt}$. Thus

$$C(t) \sim \frac{1}{9}\frac{2}{\sqrt{4\pi Dt}} = \frac{\sqrt{3}}{18\sqrt{Jt}}, \tag{25}$$

which is Eq. (7) of the main text at $J = 1$.

### 1. *Return Probabilities from Free-Fermion Determinants*

The free-fermion representation also lets us compute the time-dependent contribution of a specified wall. Consider the two product configurations related by the local spin flip. If the flip shifts the wall without changing the rods, the same rod dynamics occurs in both configurations. The spin coherence survives in this contribution when the marked wall returns to its initial physical position and the local sites needed for the reverse flip are unoccupied by rods. The wall displacement is determined by the net number of rods that have passed it. For an isolated wall, its return can therefore be expressed as a zero-transfer condition on the number of fermions to its left, together with local empty-site conditions. These conditions concern the wall contribution; the full correlator also includes the other allowed local arrangements and an average over initial spin configurations.

To evaluate such probabilities, let $h$ be the single-particle hopping matrix on the contracted open chain of $L_y$ sites, with $h_{y,y+1} = h_{y+1,y} = J/2$, and let $U_t = e^{-iht}$. For an initial occupied set $o$, the evolved many-body state is a Slater determinant: its occupied single-particle orbitals are the columns of $U_t$ indexed by $o$. Its correlation matrix is consequently

$$G_t(y, z) = \sum_{r\in o}(U_t)_{yr}(U_t)^*_{zr}. \tag{26}$$

For a set $A$ of contracted sites, define $N_A = \sum_{y\in A} n_y$, where $n_y$ is the fermion occupation. The joint generating function of $N_A$ and the occupations of two sites $a, b \notin A$ follows from the standard free-fermion determinant identity [14, 15]:

$$\left\langle e^{i\kappa N_A + i\kappa_a n_a + i\kappa_b n_b}\right\rangle_t = \det\left[\mathbb{I}_\Omega + D_\kappa\, G_t\big|_\Omega\right], \qquad \Omega = A \cup \{a, b\}, \tag{27}$$

where $D_\kappa$ is diagonal, with entries $e^{i\kappa} - 1$ on $A$, $e^{i\kappa_a} - 1$ at $a$, and $e^{i\kappa_b} - 1$ at $b$. The Fourier integral selects $N_A(t) = N_A(0)$, and the factors $1 - n_a$ and $1 - n_b$ impose the empty-site constraints. Using $1 - n = \frac{1}{2}\sum_{\kappa'\in\{0,\pi\}} e^{i\kappa' n}$ gives

$$\mathbb{P}\left[\Delta N_A = 0,\, n_a = n_b = 0\right] = \int_0^{2\pi} \frac{d\kappa}{2\pi}\, e^{-i\kappa N_A(0)}\, \frac{1}{4} \sum_{\kappa_a,\kappa_b\in\{0,\pi\}} \left\langle e^{i\kappa N_A + i\kappa_a n_a + i\kappa_b n_b}\right\rangle_t. \tag{28}$$

For the isolated-wall term, $A = \{y < y_0\}$ and the two empty sites are $a = y_0$, $b = y_0 + 1$. The neighboring contribution, in which a marked empty site represents the wall, is evaluated with that site at $y_0 - 1$ and site $y_0$ occupied. In the numerical evaluation we sample several hundred Bernoulli($\frac{1}{4}$) occupied sets on $L_y = 161$ contracted sites, conditioned on the local rod and wall arrangement, and use 32 values of $\kappa$ to discretize the Fourier integral. This calculation was benchmarked against exact diagonalization of the folded chain for $L = 20$ physical sites.

## B. Finite Anisotropy

At finite $\Delta = \cosh\eta > 1$, with $q = e^\eta$, the Bethe-ansatz quasiparticles called $s$-strings describe bound states of $s$ spin flips. Such a bound state need not form a compact spin domain in a high-temperature configuration, as discussed below. An $s$-string has an intrinsic hopping scale $\gamma_s \sim Jq^{1-s}$. Its recoil from collisions with lighter strings is diffusive.

The coherence considered in the main text changes the size of a heavy string by one. We assume that collisions preserve this coherence until the two branches undergo different intrinsic motion, giving a lifetime $\tau_s \sim \gamma_s^{-1}$. This is shorter than the time $D/v_s^2$ at which ballistic motion dominates the diffusive displacement.

Let $s_1$ be the first string size whose content differs between the branches before and after the local flip. We call the local spin configurations supporting these coherences hosts. The counting below gives their weight for $s_1 \geq s$,

$$\rho(s) = \frac{1}{(s+1)^2}. \tag{29}$$

At the folded point only magnons move, so the surviving weight is $\rho(2) = 1/9$. At finite anisotropy the cutoff grows as $s^*(t) \simeq 1 + \ln(Jt)/\ln q$. Within our kinetic framework, this yields

$$C(t) \simeq \rho(s^*(t)) P_{\mathrm{ret}}(t), \qquad P_{\mathrm{ret}}(t) \sim \frac{2}{\sqrt{4\pi D(\Delta) t}}, \tag{30}$$

which gives the $t^{-1/2}/\ln^2 t$ law in Eq. (8) of the main text. The counting fixes the asymptotic weight, but not the microscopic factor in $\tau_s$; this factor is absorbed into $t_0$. At $\Delta = \infty$, $D = 4J/(3\pi)$ [11].

### C. Strings in a Random Configuration

In real space, compact domains of length $s$ have density $2^{-(s+1)}$, whereas the thermodynamic Bethe ansatz (TBA) gives, for all $\Delta \geq 1$ [16],

$$\int d\theta\, \rho_s(\theta) = \frac{1}{s(s+1)(s+2)}, \qquad n_s = \frac{1}{(s+1)^2}, \qquad \int d\theta\, \rho_s^{\mathrm{tot}}(\theta) = \frac{s+1}{s(s+2)}. \tag{31}$$

For $s \geq 3$ there are far more strings than compact domains, because a large string at high temperature is not compact: $\uparrow\uparrow\uparrow\downarrow\uparrow\uparrow\uparrow\downarrow^5$ contains a 5-string with a 1-string passing through it. The following recursive construction reproduces the TBA densities from the statistics of a random spin configuration.

We start from i.i.d. spins and remove strings recursively. At level $s = 1, 2, 3, \ldots$, every maximal run of exactly $s$ equal spins, which we call a *minimal domain*, is an $s$-string. We remove it together with the $s$ sites immediately to its right (its partners), then re-read the remaining sequence, merging adjacent equal spins. Runs of consecutive minimal domains are consumed greedily from the left, the last one of an odd run taking its partners from the following longer domain. After level $s$ every domain has length $\geq s+1$. Each $s$-string, viewed in the sequence from which smaller strings have been removed, is a block $\downarrow^s\uparrow^s$ or $\uparrow^s\downarrow^s$. We call the sequence left after level $s$ the *level-$s$ residual*, and the size of one of its domains the *true size*.

To obtain the string densities, suppose the level-$s$ sequence has i.i.d. domain lengths $\ell \geq s$ with a geometric excess, $\mathbb{P}(\ell = s+k) = p_s(1-p_s)^k$ ($p_1 = \frac{1}{2}$ for Bernoulli spins). Write it as non-minimal domains $E_i$ with excesses $k_i \geq 1$, separated by runs of $m_i$ minimal domains, $\mathbb{P}(m_i = m) = (1-p_s)p_s^m$. At level $s$, the minimal domains in each run are removed in pairs. An odd run also removes the first $s$ sites of the following $E_i$, whose remainder merges with $E_{i-1}$. A new domain is $E_a$ followed by the remainders of $E_{a+1}, \ldots, E_b$, with the intervening runs odd and the flanking runs even; its length is $s + \sum_{i=a}^{b} k_i$ with $b-a+1$ geometric. A geometric number of i.i.d. geometric variables is geometric, with parameter $p_s/(1+p_s)$, so

$$p_{s+1} = \frac{p_s}{1+p_s} \qquad \Longrightarrow \qquad p_s = \frac{1}{s+1}. \tag{32}$$

The mean domain length at level $s$ is $2s$. The sequence has length $2/(s+1)$ per original site, giving a domain density $\nu_s = 1/[s(s+1)]$. Since a run of $m$ minimal domains yields $\lceil m/2 \rceil$ strings, we obtain $N_s = \nu_s p_s/(1+p_s) = 1/[s(s+1)(s+2)]$ and $n_s = p_s^2 = 1/(s+1)^2$, which are Eq. 31.

We can also check the entropy of this construction. The TBA vacancy entropy $L_s\mathcal{H}(n_s)$, $L_s = (s+1)/[s(s+2)]$, equals the conditional entropy of the level-$s$ sequence given the level-$(s+1)$ sequence, $\mathcal{H}(p_s)/s - \mathcal{H}(p_{s+1})/(s+1)$, and the sum over $s$ is $\ln 2$.

The construction thus reproduces the integrated densities and fillings in Eq. 31, without reference to the anisotropy. The walls of the level-$(s-1)$ residual, whose domains have true size $\geq s$, are the ends of the strings of size $\geq s$, with density $1/[s(s+1)]$, which is the counting used above.

### D. Counting the Coherent Hosts

A flip at a wall creates a coherence between the branches $A$ and $B = \sigma_x^+ A$. In the folded chain it belongs to the frozen-pattern channel if it does not create or absorb a magnon. Our finite-$\Delta$ kinetic picture applies this criterion at each level of the hierarchy. The branches evolve identically, up to the rigid displacement of the frozen pattern, as long as every mobile string of $A$ is also a string of $B$ at the same place and vice versa. The first string size $s_1$ at which the two contents differ sets the host lifetime $\tau_{s_1} = 1/\gamma_{s_1}$: from then on, one branch contains a mobile object that the other lacks. This subsection determines its distribution in the $T = \infty$ ensemble, using the decomposition of Sec. IV C.

We begin with the level-1 rule. Let the flipped site $x$ have $s_{x-1} \neq s_{x+1}$, and orient the configuration so that $s_x = s_{x+1}$: the wall sits at the bond $(x-1, x)$ and the flip moves it to $(x, x+1)$. Call $D_L$ the domain ending at $x-1$ (the *growing* domain, which the flipped spin joins) and $D_R$ the domain beginning at $x$ (the *shrinking* domain, which it leaves).

At level 1 the minimal domains are lone spins. On the left, if $D_L$ is minimal it belongs to a run of $m''$ consecutive minimal domains ending at $D_L$, which the flip removes from that run, and the number of rods $\lceil m''/2 \rceil$ is unchanged if and only if $m''$ is even. On the right, if $D_R$ has length exactly 2 it becomes minimal and joins the run of $m'$ minimal domains that follows it, and the number of rods is unchanged if and only if $m'$ is odd. If $D_L$ is not minimal, or $D_R$ has length $\geq 3$, the corresponding side is harmless.

The same reasoning applies at level $s$. If the flip is safe at levels $1, \ldots, s-1$, the level-$(s-1)$ residuals of $A$ and $B$ (Sec. IV C) have the same strings removed and differ only by the moved wall. In these residuals, $D_L$ and $D_R$ denote the domains adjacent to the wall, of lengths $\geq s$, and the minimal domains are those of length $s$.

The two conditions now decide whether an $s$-string is absorbed or created. The flip is unsafe on the left if and only if $D_L$ is minimal and the run of minimal domains ending at it has odd length. It is unsafe on the right if and only if $D_R$ has length exactly $s+1$ and the run of minimal domains following it has even length. A shrinking domain of length $s$ cannot occur: the residual of $B$ would contain a domain of length $s-1$, which the previous level would have removed, contradicting safety there. The host level $s_1$ is the first unsafe level.

This rule does not depend on which branch moves first: a shrinking domain of length $s+1$ followed by an odd run is safe because in $B$ the now-minimal domain pairs with the first member of that run into the same $s$-string, at the same place, that the run's first two members formed in $A$, by the identification of Sec. IV A one level up.

We can now determine the distribution of host levels. At level $s$ the residual has i.i.d. geometric domain lengths, $\mathbb{P}(\ell = s+k) = p_s(1-p_s)^k$ with $p_s = 1/(s+1)$ (Eq. 32), and memorylessness makes the two sides independent of each other and of the lower levels. On the left, $D_L$ is minimal with probability $p_s$, and the run ending at it has $m'' = 1 +$ (geometric) members, $\mathbb{P}(m'' = m) = (1-p_s)p_s^{m-1}$, so $\mathbb{P}(m'' \text{ odd}) = 1/(1+p_s)$. On the right, $D_R$ has length $s+1$ with probability $p_s$ and $\mathbb{P}(m' \text{ even}) = 1/(1+p_s)$ likewise. Each side is therefore unsafe with probability $p_s/(1+p_s)$. The probability of passing the level is $[1/(1+p_s)]^2 = [(s+1)/(s+2)]^2$, and

$$\mathbb{P}\big[s_1 \geq s \,\big|\, \text{flip at a wall}\big] = \prod_{s'=1}^{s-1} \left(\frac{s'+1}{s'+2}\right)^2 = \frac{4}{(s+1)^2} = 4n_s, \qquad \rho(s) \equiv \mathbb{P}\big[s_1 \geq s\big] = \frac{1}{4} \cdot \frac{4}{(s+1)^2} = \frac{1}{(s+1)^2}, \tag{33}$$

where $\frac{1}{4} = \mathbb{P}[s_x = \downarrow,\ s_{x-1} \neq s_{x+1}]$ and the product telescopes because $1/(1+p_s) = p_{s+1}/p_s$, the recursion of Eq. 32.

To assemble the correlator, define the host weights $a_s = \rho(s) - \rho(s+1) = (2s+3)/[(s+1)^2(s+2)^2]$, distinct from the TBA rapidity density $\rho_s(\theta)$. We approximate the survival of a level-$s$ host by $w_s(t) = \theta(\tau_s - t)$, and use the same heavy-string return probability $P_{\text{ret}}$ for the wall in all surviving hosts. This gives

$$C(t) \simeq P_{\text{ret}}(t) \sum_{s \geq 2} a_s\, w_s(t) = \rho\big(s^*(t)\big)\, P_{\text{ret}}(t), \qquad s^*(t) = \max\left[2,\ 1 + \frac{\ln Jt}{\ln q}\right], \tag{34}$$

which is Eq. 30. The hierarchy does not fix the $O(1)$ factor in the coherent-hop time $\tau_s$, and hence does not determine $s^*$ precisely. This uncertainty shifts $s^*$ by a constant and the prefactor by $1 + O(1/s^*)$.

## V. FINITE MAGNETIZATION

We now consider how a nonzero magnetization changes the coherent-host weight and the quantum-group overlap. The biased infinite-temperature ensemble is the product state $\rho_m \propto \prod_j \left[p\, n_j^{\uparrow} + (1-p)\, n_j^{\downarrow}\right]$ with up-probability $p = \frac{1}{2}(1+m)$, $\langle \sigma_j^z \rangle = m$. It is a function of $S^z$, hence stationary under every Hamiltonian considered here, and the correlator is $C_m(t) = \mathrm{Tr}[\rho_m\, \sigma_x^-(t)\sigma_x^+]$, with $C_m(0) = \mathbb{P}[s_x = \downarrow] = 1 - p$.

Spin flip combined with reflection maps $m \to -m$ and leaves every chain invariant, so we take $m \geq 0$. Throughout this section

$$n \equiv p(1-p) = \frac{1-m^2}{4}, \qquad 1-n = \frac{3+m^2}{4}, \qquad 1-2n = \frac{1+m^2}{2}. \tag{35}$$

### A. Folded Chain

For Bernoulli($p$) spins, the folded-chain calculation proceeds as at half filling, with modified magnon and frozen-pattern densities. Evaluating the heavy-string diffusion formulas of Ref. [17] in the folded limit gives (see also [13])

$$D(m) = \frac{4}{\pi}\frac{1-m^2}{3+m^2}, \tag{36}$$

which reduces to the infinite-temperature expression for $m = 0$.

The coherent host weight becomes

$$\rho(m) = \frac{1-m}{2}\,\frac{2n^2}{(1-n)^2}\Big|_{n=\frac{1-m^2}{4}} = \frac{(1-m)(1-m^2)^2}{(3+m^2)^2}. \tag{37}$$

Multiplying this weight by $2/\sqrt{4\pi D(m)t}$ gives the long-time folded correlator.

### B. String Densities at Finite Magnetization

At finite $\Delta > 1$, the hosts surviving at time $t$ are the flips of host level $s_1 \geq s^*(t)$, $s^* \simeq 1 + \ln(Jt)/\ln q$ (Secs. IV B and IV D). At $p = \frac{1}{2}$ their weight is $1/(s^*+1)^2$ and the suppression is logarithmic. At finite magnetization the infinite-temperature TBA has a field: the string fillings become [16]

$$n_s = \frac{\sinh^2\mu}{\sinh^2[(s+1)\mu]}, \qquad \tanh\mu = m, \tag{38}$$

which reduce to $1/(s+1)^2$ at $\mu \to 0$ and decay as $e^{-2\mu s}$ at large $s$. The densities $\rho_s = n_s\rho_s^{\mathrm{tot}}$ also decay exponentially. The weight of hosts of level $\geq s^*$ is therefore $\propto e^{-2\mu s^*} \sim (t)^{-2\mu/\ln q}$, and

$$C_m(t) \propto t^{-1/2-2\mu/\ln q}, \qquad \mu = \operatorname{artanh} m. \tag{39}$$

The $t^{-1/2}$ factor comes from the return probability of the (temporarily) frozen pattern, and the continuously varying power law comes from balancing the exponentially decaying probabilities and exponentially growing lifetimes of large strings [17].

### C. Mazur Bound at Finite Magnetization

For the stationary biased ensemble, we use $(A, B)_m = \mathrm{Tr}(\rho_m A^\dagger B)$. The time-average projection argument still applies. In the $q \to \infty$ limit its static overlap is

$$\langle \sigma_x^- \tilde{e}\rangle_m = \mathbb{P}_p[d(c) = x]. \tag{40}$$

The finite-$q$ inequality of Sec. II was established for the full unbiased trace and is not invoked for this weighted trace.

For $m \neq 0$, the path has a drift and its global minimum is concentrated near one edge. A bulk site's class weight is exponentially small in its distance from that edge, with a crossover length of order $m^{-2}$ at small bias. Sites near the favored edge can retain a finite overlap, but their distance to the window boundary does not grow with its size. Thus this finite-size Mazur construction does not yield the algebraic bulk lower bound of Sec. III at fixed nonzero magnetization.

## VI. NUMERICAL METHODS

### A. Dynamical Typicality

We compute infinite-temperature correlators using dynamical typicality [18, 19]. For the XXZ, folded and Fredkin chains, correlators on $L \leq 22$ were computed as $\langle O_1 O_2(t)\rangle \simeq \langle\psi|O_1 e^{iHt} O_2 e^{-iHt}|\psi\rangle/\langle\psi|\psi\rangle$ for Gaussian random $|\psi\rangle$ in the full Hilbert space, with statistical error of order $2^{-L/2}$ per vector. Sector- and class-resolved correlators are obtained by inserting projectors before $\sigma_x^+$. Time evolution used a Chebyshev expansion in steps $\tau = 0.5$ with spectral bounds from a 40-step Lanczos iteration, or a Krylov exponential in the magnetization sector. The folded-chain profiles used 8 vectors on rings of $L = 16$ to $t = 24$, 4 on $L = 18$ to $t = 20$ and 3 on $L = 20$ to $t = 10$; the sector-resolved correlators used open chains of $L = 20$ with 6 vectors (frozen-wall, magnon and transfer sectors) and 8 vectors (frozen-wall sector at $\Delta = 3, 4, 6, \infty$ and string correlator); error bars are standard errors over vectors, and the fitted decay rates carry jackknife errors. This approach was benchmarked against exact diagonalization for $L = 20$ sites.

### B. Long-Chain Folded Data and Finite Anisotropy

The folded correlator in Fig. 3 of the main text was computed on open chains of $L = 256, 512, 1024$ by unbiased Monte Carlo sampling of free-fermion Slater determinants, using up to $10^6$ samples [10, 20, 21]; the return constraints and determinant formula are explained in Sec. IV A 1. The finite-$\Delta$ data at $\Delta = 10, 5$ use dynamical typicality on periodic chains of $L = 24$ [18, 19]. The slow part is extracted by fitting

$$C(t) = C_0(t) + C_1(t)\cos\Delta t + S_1(t)\sin\Delta t, \tag{41}$$

in a sliding window. Fitting $\sqrt{t}\,C_0(t) = a/\ln^2(t/t_0)$ over $12 \leq Jt \leq 29$ gives $(t_0, a) = (0.11, 2.39)$ at $\Delta = 10$ and $(0.76, 0.76)$ at $\Delta = 5$.

### C. Mazur Bounds and Plateaus

Every magnetization sector of an open chain of $L \leq 16$ sites was diagonalized in full; for $L = 15, 16$ the sectors with more than $L/2$ up spins were obtained from those with fewer by spin flip combined with reflection, a symmetry of all four chains (checked on $H$ to machine precision). For the static overlap, we choose joint eigenvectors of $H$ and $E^\dagger E$ in each magnetization sector; inside groups of levels degenerate to $10^{-7}$ we first diagonalized $E^\dagger E$ and then, within each of its eigenspaces, $H$. Then $\mathcal{R}v_n = Ev_n/|Ev_n|$ (and 0 if $Ev_n = 0$); the values $|Ev_n|^2$ reproduce Eq. 4 and their number the rank implied by Eq. 16 in every case. $\langle\sigma_x^- \mathcal{R}\rangle = 2^{-L}\sum_n \langle v_n|\sigma_x^- \mathcal{R}|v_n\rangle$, which is basis independent within each component.

The plateau is obtained by summing matrix elements between degenerate states. The contribution from symmetry-related pairs is $2^{-L}\sum_n |\langle\mathcal{R}v_n|\sigma_x^+|v_n\rangle|^2$, the sum over the exactly degenerate pairs $|j, m\rangle, |j, m+1\rangle$ of the $U_q(sl_2)$ multiplets; this estimate coincides with the sum over all pairs of eigenstates of adjacent sectors with $|\varepsilon_n - \varepsilon_{n'}| < 10^{-10}$ to a relative accuracy of $10^{-4}$ on every chain with $L \leq 15$, and of 1% at $L = 16$, $q = 10$, where quasi-degeneracies extend below $10^{-10}$. A looser matching, $10^{-7}$, also counts quasi-degenerate pairs split by the large-$q$ hierarchy, and overstates $\overline{C}$ at $q = 10$ by 3.5% on $L = 14$ and 8% on $L = 16$. In the Fredkin chain, which has no such structure, $\overline{C}$ is the sum over pairs with $|\varepsilon_n - \varepsilon_{n'}| < 10^{-10}$; it is unchanged for tolerances between $10^{-12}$ and $10^{-7}$ (and changes by less than 0.3% up to $10^{-5}$).

Using the same eigenstates, $C(t)$ and its running average are the sums over all pairs with $\cos[(\varepsilon_{n'} - \varepsilon_n)t]$ and $\sin[(\varepsilon_{n'} - \varepsilon_n)T]/(\varepsilon_{n'} - \varepsilon_n)T$ inserted, evaluated on a grid of 3200 times in $0 \leq t \leq 10^5$. The class weights $\mathbb{P}[d(c) = x]$ were obtained by exhaustive enumeration and agree with Eq. 10. For $L \leq 14$ the bounds, and the plateaus with $10^{-7}$ matching, agree to all digits quoted with an independent computation that obtains $\mathcal{R}$ from the singular-value decomposition of $E$.

### D. Finite Magnetization

For the biased ensemble, we sample typicality states of the form $|\psi\rangle = \sum_c \sqrt{\rho_m(c)}\, z_c|c\rangle$, $z_c$ complex Gaussian, so that $\langle\psi|\sigma_x^-(t)\sigma_x^+|\psi\rangle/\langle\psi|\psi\rangle$ estimates $C_m(t)$. Sector or class projections are implemented by restricting the sum over $c$. These calculations use rings of $L = 16$–$22$ with 2–10 vectors and Krylov time steps of 0.25 (folded, XXZ) or 0.5

(Fredkin). We also use full diagonalization in magnetization sectors on $L \leq 16$ for the Fredkin chain at $p \geq 0.85$. The class-resolved open-chain data for the Fredkin chain evolve individual product configurations by Krylov exponentials in the two magnetization sectors of the branches ($L \leq 22$). The class weights $\mathbb{P}_p[d(c) = x]$ were computed by a biased-walk height recursion and checked by enumeration on $L \leq 12$.

### Appendix A: Proof of the overlap inequality for all $q \geq 1$

This appendix proves Eq. (15), $\langle \sigma_x^- \mathcal{R} \rangle \geq p_{L,x}$ for every $L \geq 1$, every site $x$ and every $q \geq 1$. As stated in Sec. II C, the argument was generated with the assistance of large language models with minimal human input, and has been formalized and machine-checked in Lean 4 (Mathlib): the statement $\langle \sigma_x^- \mathcal{R} \rangle \geq p_{L,x}$ for all $L$, $x$ and $q \geq 1$, with $E$, the canonical polar factor and the trace defined exactly as in Eqs. (3), (5) and (10), is proved without additional axioms or unproved hypotheses [22]. We have verified its intermediate statements numerically on random instances and its conclusion against the exact overlaps of Sec. II C. This proof is included for the sake of completeness, but for our purposes, the basic intuition is spelled out in Sec. II A.

To establish the bound for arbitrary $q \geq 1$, we decompose the spins on either side of $x$ into quantum-group multiplets $V_a$ and $V_b$. The trace becomes a sum, with positive multiplicities, of overlaps on $V_a \otimes V_{1/2} \otimes V_b$. The $q \to \infty$ overlap in each block is one, so it suffices to bound the finite-$q$ overlap by one as well. The main step is to compare the polar factors of the raising operators on two multiplets. These raising blocks are nonnegative bidiagonal matrices. Their polar factors can be represented as zero-temperature transverse magnetizations of an Ising chain. Ising duality and ferromagnetic correlation inequalities then give a comparison between reduced raising amplitudes. Finally, coupling $V_a$ to the local spin gives two channels, $a \pm 1/2$. Their contributions have opposite signs, but the positive channel has the larger reduced amplitude. This controls their difference and proves the block bound.

#### 1. Bidiagonal Polar Factors and Ising Correlations

We use the standard ferromagnetic correlation inequality: in a classical Ising model with nonnegative interactions, every spin-product correlation is nondecreasing in each coupling [23]. The Trotter representation extends this to order correlations of

$$\mathcal{K} = -\sum_A J_A Z_A - \sum_i h_i X_i, \qquad J_A, h_i \geq 0: \tag{A1}$$

at fixed transverse fields, $\langle Z_B \rangle_{\mathcal{K},\beta_{\mathrm{I}}}$ is nondecreasing in every $J_A$. We apply this standard result to the dual Ising chain below. Let $A(d, b)$ be a real square lower bidiagonal matrix with diagonal entries $d_i \geq 0$ and subdiagonal entries $b_i \geq 0$. If $d_i' \geq d_i$ for all $i$, with $b$ fixed, then

$$[\mathrm{Pol}(A(d', b))]_{ii} \geq [\mathrm{Pol}(A(d, b))]_{ii} \quad \text{for every } i. \tag{A2}$$

The same result holds for upper bidiagonal matrices by transposition, including singular matrices. To establish this inequality, consider an open Ising chain of $N$ spins,

$$\mathcal{K}_A = -\sum_{i=1}^{N} d_i X_i - \sum_{i=1}^{N-1} b_i Z_i Z_{i+1}. \tag{A3}$$

Define Majorana operators

$$\alpha_i = (-1)^{i-1} \left( \prod_{k<i} X_k \right) Z_i, \qquad \gamma_i = (-1)^{i-1} \left( \prod_{k<i} X_k \right) Y_i. \tag{A4}$$

They satisfy the canonical anticommutation relations, $X_i = i\alpha_i\gamma_i$, and direct multiplication gives

$$\mathcal{K}_A = -i \sum_{r,s} \alpha_r A(d, b)_{rs} \gamma_s. \tag{A5}$$

For a real singular-value decomposition $A = U \operatorname{diag}(s_\ell) V^T$, orthogonal rotations of the two Majorana families turn (A5) into independent terms $-s_\ell(i\widehat{\alpha}_\ell\widehat{\gamma}_\ell)$. It follows that

$$\langle X_i \rangle_{\mathcal{K}_A,\beta_{\mathrm{I}}} = [U \operatorname{diag}(\tanh(\beta_{\mathrm{I}} s_\ell)) V^T]_{ii}. \tag{A6}$$

This identity also holds at $s_\ell = 0$, where the thermal contribution is zero. Now perform the exact Pauli change of variables

$$\widetilde{Z}_i = \prod_{k=1}^{i} X_k, \qquad \widetilde{X}_i = Z_i Z_{i+1}\ (i < N), \qquad \widetilde{X}_N = Z_N, \qquad \widetilde{Z}_0 = \mathbb{I}. \tag{A7}$$

These operators form $N$ independent Pauli pairs. In the dual variables,

$$\mathcal{K}_A = -d_1 \widetilde{Z}_1 - \sum_{i=2}^{N} d_i \widetilde{Z}_{i-1}\widetilde{Z}_i - \sum_{i=1}^{N-1} b_i \widetilde{X}_i, \qquad X_i = \widetilde{Z}_{i-1}\widetilde{Z}_i. \tag{A8}$$

Thus each $d_i$ is a nonnegative order coupling (a singleton field for $i = 1$), while the $b_i$ are fixed transverse fields. The ferromagnetic correlation inequality implies that every expectation in (A6) is nondecreasing in every $d_j$. Finally, take $\beta_{\mathrm{I}} \to \infty$. The matrix in (A6) tends to the canonical polar factor: each positive singular value contributes one, and each zero singular value contributes zero. This proves (A2) without using continuity of the polar factor at a rank change.

### 2. Comparison of Reduced Quantum-Group Ladders

For $j > 0$ define coefficients $f^{(b)}_{j,m}$ by

$$\mathrm{Tr}_{V_b} \mathrm{Pol}(E_{j,b}) = \sum_{m=-j}^{j-1} f^{(b)}_{j,m} |j, m+1\rangle\langle j, m|. \tag{A9}$$

Only the displayed weight transitions occur, because $E_{j,b}$ raises the total weight by one and its canonical polar factor has the same weight property. For every allowed $m$, $q \geq 1$, and $b \geq 0$,

$$f^{(b)}_{j+1,m} \geq f^{(b)}_{j,m}, \qquad f^{(b)}_{j,m} \geq f^{(b)}_{1/2,-1/2}. \tag{A10}$$

To compare the reduced amplitudes, extend $\lambda_j(m)$ in (4) by zero outside $-j \leq m < j$. Fix a source total weight $M$ and order the right weights as $n = -b, -b+1, \ldots, b$. Define a square auxiliary matrix $A_{j,M}$ by

$$(A_{j,M})_{n,n} = q^n \lambda_j(M-n), \qquad (A_{j,M})_{n+1,n} = q^{-(M-n)} \lambda_b(n). \tag{A11}$$

It is a nonnegative lower bidiagonal matrix. The physical raising block can be rectangular, but its polar entries are contained in this square matrix. Its physical source columns satisfy $M - j \leq n \leq M + j$, and its physical target rows satisfy $M + 1 - j \leq n \leq M + 1 + j$, intersected with $[-b, b]$. A nonzero diagonal edge is physical precisely when $M - j + 1 \leq n \leq M + j$. A subdiagonal edge from column $n$ to row $n+1$ is physical precisely when $M - j \leq n \leq M + j$. Every installed subdiagonal edge outside this interval joins two nonphysical vertices, and the adjacent diagonal edges vanish. Hence the bipartite graph of $A_{j,M}$ separates into the physical component, isolated auxiliary subdiagonal dimers, and possibly isolated vertices. After separate row and column permutations it is an orthogonal direct sum. Canonical polar factors respect such sums, so for a fixed physical transition $m \to m+1$,

$$f^{(b)}_{j,m} = \sum_{n=-b}^{b} [\mathrm{Pol}(A_{j,m+n})]_{n,n}. \tag{A12}$$

The auxiliary dimers have zero diagonal polar entries. For the first comparison, the subdiagonal in (A11) does not depend on $j$. Increasing $j$ to $j+1$ increases every zero-extended $\lambda_j(m)$, including previously forbidden transitions. On the common allowed interval this follows from

$$\lambda_j(m)^2 = [j]_q [j+1]_q - [m]_q [m+1]_q. \tag{A13}$$

The monotonicity inequality (A2), applied separately at each $M = m + n$, proves the first inequality in (A10).

For the second comparison, retain only the transition $m \to m+1$ in the left ladder, setting all other left raising amplitudes to zero. At each total weight this decreases only diagonal entries in (A11); the subdiagonal is unchanged. The selected polar entries can therefore only decrease. Their physical component is now a two-level left subsystem

with weights $m, m+1$; the other components again have zero diagonal polar entries. Put $\mu = m + 1/2$. On this two-level subsystem, $K = q^{\mu} K_{1/2}$ and the raising operator is $\lambda_j(m)\sigma^+$. Its coproduct with $V_b$ is

$$q^{-\mu}\big(g\,\sigma^+ \otimes K_b + K_{1/2}^{-1} \otimes E_b\big), \qquad g = q^{m+1/2}\lambda_j(m). \tag{A14}$$

The positive overall scalar does not change the polar factor. For positive integers $r$, $[r]_q \geq q^{r-1}$ when $q \geq 1$. Both $j-m$ and $j+m+1$ are positive integers, so

$$g^2 = q^{2m+1}[j-m]_q[j+m+1]_q \geq q^{2(j+m)} \geq 1. \tag{A15}$$

In the two-level raising blocks, increasing $g$ from one increases only the diagonal entries. Another application of the monotonicity inequality (A2) compares (A14) with the standard $V_{1/2} \otimes V_b$ coproduct. Summing its selected entries proves the second inequality in (A10).

### 3. Overlap on Three Quantum-Group Multiplets

For $a, b \geq 0$, define the unnormalized middle-spin overlap

$$\mathcal{O}_{a,b}(q) = \mathrm{Tr}_{V_a \otimes V_{1/2} \otimes V_b}\left[(\mathbb{I}_a \otimes \sigma^- \otimes \mathbb{I}_b)\,\mathrm{Pol}(E_{a,1/2,b})\right], \tag{A16}$$

where the three-factor coproduct is obtained by iterating (2). Set $r = 2a+1$, $\varrho = q^{-1}$ and, for $0 \leq k \leq r$, let

$$u_k = \sqrt{\frac{1-\varrho^{2k}}{1-\varrho^{2r}}}, \qquad v_k = \varrho^k\sqrt{\frac{1-\varrho^{2(r-k)}}{1-\varrho^{2r}}}. \tag{A17}$$

At $q = 1$ these mean $u_k = \sqrt{k/r}$ and $v_k = \sqrt{(r-k)/r}$. They satisfy $u_k^2 + v_k^2 = 1$. At total weight $M_k = k - a - 1/2$, use the ordered product basis

$$\big(|a, M_k - 1/2\rangle \otimes |\uparrow\rangle, \quad |a, M_k + 1/2\rangle \otimes |\downarrow\rangle\big). \tag{A18}$$

The upper-spin $J_+ = a + 1/2$ vector is $(u_k, v_k)$, and the lower-spin $J_- = a - 1/2$ vector is $(-v_k, u_k)$. The latter exists only for $1 \leq k \leq r-1$, and is absent altogether if $a = 0$. Endpoint product vectors with an out-of-range left weight are omitted. Direct substitution in (2) verifies that these orthonormal columns intertwine the two spin ladders: the upper coefficient is $\sqrt{[r-k]_q[k+1]_q}$ and the lower coefficient is $\sqrt{[r-k-1]_q[k]_q}$.

Write $u_k = \cos\theta_k$, $v_k = \sin\theta_k$. Since $u_k$ increases, the angles decrease from $\theta_0 = \pi/2$ to $\theta_r = 0$. Define

$$U_k = u_{k+1}v_k, \qquad V_k = v_{k+1}u_k. \tag{A19}$$

Then $U_k \geq V_k \geq 0$ and

$$U_k - V_k = \sin(\theta_k - \theta_{k+1}), \qquad \sum_{k=0}^{r-1}(U_k - V_k) \geq 1. \tag{A20}$$

For the last inequality, the nonnegative increments sum to $\pi/2$, and repeated application of $\sin s + \sin t \geq \sin(s+t)$ for $s, t \geq 0$, $s + t \leq \pi/2$, gives the result.

In $V_a \otimes V_{1/2}$, the polar ladder has unit coefficient within each multiplet. The local half-spin raising transition has upper-channel coefficient $U_k$ and lower-channel coefficient $-V_k$, as follows immediately from the two columns in (A18). Hence

$$\mathcal{O}_{a,0}(q) = \sum_{k=0}^{r-1}(U_k - V_k) \geq 1. \tag{A21}$$

Reversing the tensor order and reversing both spin weights sends $E_{1/2,b}$ to $E_{b,1/2}^{\dagger}$. It sends the local lowering operator to the local raising operator. Canonical polar factors respect adjoints and unitary conjugations, and all matrices are real, so

$$f^{(b)}_{1/2,-1/2} = \mathcal{O}_{0,b}(q) = \mathcal{O}_{b,0}(q) \geq 1. \tag{A22}$$

For all half-integers $a, b \geq 0$ and all $q \geq 1$,

$$\mathcal{O}_{a,b}(q) \geq \mathcal{O}_{a,0}(q)\mathcal{O}_{0,b}(q) \geq 1. \tag{A23}$$

To see this, first couple $V_a \otimes V_{1/2}$ into $J_+$ and $J_-$. In this unitary basis, the three-factor raising generator is the orthogonal sum of $E_{J_+,b}$ and $E_{J_-,b}$. Its canonical polar factor has the same decomposition. After tracing over $V_b$, (A9) therefore gives the exact identity

$$\mathcal{O}_{a,b}(q) = \sum_{k=0}^{r-1} \left[ f^{(b)}_{a+1/2,M_k} U_k - f^{(b)}_{a-1/2,M_k} V_k \right]. \tag{A24}$$

At $k = 0, r-1$, the lower ladder is absent, but $V_k = 0$, so the corresponding term is simply omitted. For an interior transition, the comparisons in Eq. (A10) and (A22) imply

$$f_+ \geq f_- \geq \mathcal{O}_{0,b}(q) \geq 1. \tag{A25}$$

Thus its summand satisfies

$$f_+U_k - f_-V_k = f_-(U_k - V_k) + (f_+ - f_-)U_k \geq \mathcal{O}_{0,b}(q)(U_k - V_k). \tag{A26}$$

For the endpoint transitions, $f_+ \geq \mathcal{O}_{0,b}(q)$ and $V_k = 0$ yield the same inequality. Sum over $k$ and use (A21) to obtain (A23).

### 4. Full-Chain Trace and Mazur Bound

We can now sum the block inequalities to obtain Eq. (15). To do so, we decompose the spins on either side of $x$ into quantum-group multiplets. The Clebsch–Gordan coefficients in Eq. A17 give

$$\mathcal{H}_{x-1} = \bigoplus_a \mathbb{C}^{m_a^{(x-1)}} \otimes V_a, \qquad \mathcal{H}_{L-x} = \bigoplus_b \mathbb{C}^{m_b^{(L-x)}} \otimes V_b. \tag{A27}$$

The ordered coproduct and its canonical polar factor act identically on the multiplicity factors. Since the local spin acts only on the middle $V_{1/2}$,

$$\mathrm{Tr}(\sigma_x^- \mathcal{R}) = \sum_{a,b} m_a^{(x-1)} m_b^{(L-x)} \mathcal{O}_{a,b}(q) \geq \left( \sum_a m_a^{(x-1)} \right) \left( \sum_b m_b^{(L-x)} \right). \tag{A28}$$

The standard spin-$\frac{1}{2}$ multiplicities are [1]

$$m_j^{(n)} = \binom{n}{n/2 - j} - \binom{n}{n/2 - j - 1}, \qquad \sum_j m_j^{(n)} = \binom{n}{\lfloor n/2 \rfloor}, \tag{A29}$$

with out-of-range binomial coefficients zero. Equation (A28) now gives the lower bound stated in (15). The product of binomial coefficients in Eq. 10 also counts the overlap with the limiting raising operator: before its first minimum the reversed walk is strictly positive, while after it the walk is weakly nonnegative. This identifies the lower bound with $p_{L,x}$ and completes the proof.



\end{document}